\documentclass[twocolumn,trackchanges]{aastex7}

\usepackage{amsmath}
\usepackage{amssymb}
\usepackage{graphicx}
\let\tablenum\relax
\usepackage{siunitx}
\usepackage{bold-extra}
\usepackage[utf8]{inputenc}

\DeclareRobustCommand{\ion}[2]{\textup{#1\,\textsc{\lowercase{#2}}}}
\DeclareUnicodeCharacter{2060}{\nolinebreak}

\newcommand\fei{\ion{Fe}{i}}
\newcommand\feii{\ion{Fe}{ii}}
\newcommand\ndi{\ion{Nd}{i}}
\newcommand\ndii{\ion{Nd}{ii}}
\newcommand\ndiii{\ion{Nd}{iii}}
\newcommand{\kms}{km\,s$^{-1}$}
\newcommand{\teff}{$T_{\rm eff}$}
\newcommand{\logg}{$\log g$}
\newcommand{\vt}{$\xi_{t}$}

\newcommand{\rproc}{$r$-process}
\newcommand{\sproc}{$s$-process}
\newcommand{\sh}{$S_\text{H}$}

\newcommand{\AB}[2]{$\mbox{[#1/#2]}$}
\newcommand{\feh}{\AB{Fe}{H}}

\newcommand{\rII}{$r$-II}

\newcommand{\logeps}{$\mathrm{A}(\text{Nd})$}
\newcommand{\logepsLTE}{$\mathrm{A}(\text{Nd})_\text{LTE}$}
\newcommand{\logepsNLTE}{$\mathrm{A}(\text{Nd})_\text{NLTE}$}
\newcommand{\avgNLTE}{$\langle\mathrm{3D}\rangle$}
\newcommand{\dnlte}{$\Delta \mathrm{A}_\text{NLTE}$}
\newcommand{\citenp}[1]{\citeauthor{#1} \citeyear{#1}}

\received{May 7, 2025}
\revised{September 19, 2025}
\accepted{September 25, 2025}

\begin{document}

\title{Investigating non-LTE abundances of Neodymium (Nd) in metal-poor FGK stars}
\shorttitle{Investigating non-LTE abundances of neodymium (Nd) in metal-poor FGK stars}
\shortauthors{Dixon, J. et al.}

\author[orcid=0000-0001-6168-3130,sname='Dixon']{John D. Dixon}
\affiliation{Department of Physics and Astronomy, Texas A\&M University, College Station, TX 77843, USA}
\affiliation{Department of Astronomy, University of Florida, Gainesville, FL 32601, USA} 
\email{johndixon@tamu.edu}

\author[orcid=0000-0002-8504-8470,sname='Ezzeddine']{Rana Ezzeddine}
\affiliation{Department of Astronomy, University of Florida, Gainesville, FL 32601, USA}  
\email{rezzeddine@ufl.edu}

\author[orcid=0000-0002-9953-7929,sname='Li']{Yangyang Li}
\affiliation{Department of Astronomy, University of Florida, Gainesville, FL 32601, USA}  
\email{yangyangli@ufl.edu}

\author[orcid=0000-0001-8253-1603,sname='Merle']{Thibault Merle}
\affiliation{Institut d'Astronomie et d'Astrophysique and Brussels Laboratory of the Universe (BLU-ULB), Universit\'e Libre de Bruxelles, CP 226, 1050 Brussels, Belgium}
\affiliation{Royal Observatory of Belgium, Avenue Circulaire 3, 1180 Brussels, Belgium}  
\email{tmerle@ulb.ac.be}

\author[orcid=0000-0001-6837-3055,sname='Bautista']{Manuel Bautista}
\affiliation{United States Department of Energy, Washington, DC 20585, USA}  
\email{manuel.bautista@science.doe.gov}

\author[orcid=0000-0001-9989-9834,sname='Guo']{Yanjun Guo}
\affiliation{Yunnan Observatories, Chinese Academy of Sciences (CAS), Kunming 650216, Yunnan, China}
\email{guoyanjun@ynao.ac.cn}

\correspondingauthor{John D. Dixon}
\email{johndixon@tamu.edu}


\begin{abstract}

The dominant site(s) of the $r$-process are a subject of current debate. Ejecta from $r$-process enrichment events like kilonovae are difficult to directly measure, so we must instead probe abundances in metal-poor stars to constrain $r$-process models. This requires state-of-the-art Non-Local Thermodynamic Equilibrium (NLTE) modeling, as LTE is a poor approximation for the low-opacity atmospheres of metal-poor giants. Neodymium (Nd) is a prominent $r$-process element detected in both near-infrared kilonovae spectra and spectra of metal-poor stars, so precise Nd stellar abundances are particularly needed to model kilonovae and constrain $r$-process sites. We thus constructed a \ndi/\ndii\ model atom to compute NLTE abundances in FGK metal-poor stars. 
We obtain $\mathrm{A}(\text{Nd})_\odot = 1.44\pm0.05$, in agreement with the meteoritic value, when calibrating the model atom with a Drawin hydrogen collision factor of $S_H=0.1$.
For a sample of metal-poor \rproc\ enhanced stars with observed optical and near-infrared \ndii\ lines, we find NLTE Nd corrections in the range $-0.3$ to $0.3$ dex. Optical and UV lines have positive NLTE corrections, whereas H band lines have negative corrections. Additionally, we compute a large grid of NLTE corrections for 122 \ndii\ spectral lines ranging from the UV to the H band, for stellar parameters of typical metal-poor FGK dwarfs and giants with $-3.00\le\mbox{[Fe/H]}\le-1.00$ and $-2.0\le\mathrm{A}(\text{Nd})\le2.0$. Within this grid, we find NLTE corrections ranging from $-0.3$ to $+0.5$ dex. Deviations from LTE are found to be strongest for blue lines with low excitation potentials in the most metal-poor giants.

\end{abstract}


\section{Introduction}\label{intro}
The origins of heavy elements are a key focus of stellar archaeology, and approximately half of all isotopes heavier than iron are typically created via the rapid neutron capture process (\rproc), which requires extreme neutron densities above ${\sim} 10^{22}$ $\mathrm{cm}^{-3}$ \citep{frebel-rev2018}. The sites of the \rproc\ are still highly debated, with collapsars \citep{siegel2019}, exotic supernovae \citep{nishimura2017}, and neutron star mergers (NSMs) \citep{watson2019} among the possible candidates. The detection of GW170817 led to the discovery of heavy element ejecta from a kilonova explosion via radioactive decay as imprinted on its spectra, prompting more research into the rates at which NSMs form the \rproc\ abundances observed in stars today. As we are unable to measure abundances from direct observations of kilonovae due to heavily blurred spectral lines from the extreme Doppler shift, models of \rproc\ events can be otherwise constrained using traces of \rproc\ elements found in second-generation metal-poor (\feh\ $< -1.0$) stars. These stars are pristine and are believed to have formed from the ejecta of one or few \rproc\ events, preserving clear signatures of \rproc\ elements in their atmospheres after birth \citep{frebel2010, frebel-rev2018}.

Determining accurate ejecta of \rproc\ events as preserved in metal-poor stars requires high precision \rproc\ elemental abundance calculations. Due to recent efforts led by large spectroscopic surveys including the R-Process Alliance \citep{rpa1, rpa2, rpa3, rpa4, rpa5}, LAMOST \citep{lamost}, GALAH \citep{desilva2015, galah1, galah2, galah3, galah4}, as well as others, the number of discovered \rproc\ enhanced stars has vastly increased in recent years. The bulk of abundance analyses of these stars have been done under the atmospheric assumption of Local Thermodynamic Equilibrium (LTE). 
Spectral lines forming in metal-poor star atmospheres, however, can significantly deviate from LTE models, particularly due to a deficiency of metal electron donors in the stellar atmosphere. This leads to a larger temperature gradient that increases over-ionization, as well as lower atmospheric opacities that allow photons to carry non-local information during spectral line formation, causing systematic (and sometimes significant) under- or over-estimation of abundances relative to LTE \citep{mashonkina2011, bergemann2014, ezzeddine2017, lind2024}. NLTE effects for spectral lines are relative to the elemental abundance, and heightened in stellar atmospheres with lower metal abundances. Therefore, to reliably calculate \rproc\ abundances in metal-poor stars, we must account for NLTE modeling effects.

Neodymium (Nd, $Z=60$) is a neutron capture element produced by both the \sproc\ and \rproc\ in similar quantities \citep{arlandini1999}, which limits its ability to act as a tracer for either process. However, according to recently generated synthetic spectra and NLTE models of kilonovae \citep{Hotokezaka2021,pognan2023}, Nd is one of the dominant lanthanides detectable in the ejecta of a kilonova, and so accurate measurements of Nd in stars known to have \rproc\ enhancement are still crucial for probing NSMs and constraining \rproc\ models. Nd has been measured in both optical and near-infrared (NIR) stellar spectra \citep{Hasselquist_2016, Afşar_2018} and it is very useful for detecting \rproc\ enrichment signatures of stars in the Galactic plane due to having several observable spectral lines in the NIR which are less affected by dust extinction \citep{salessilva2024}.
Since only lines from the dominant species (\ndii) have been measured so far in FGK stars, the abundance analysis of Nd has typically assumed LTE. \citet{abdelkawy2017} investigated NLTE abundances for Nd II lines in the Sun and found negligible average NLTE abundance corrections of ${\sim}0.01$\,dex to the standard LTE abundance. Another study by \citet{mashonkina2005} determined NLTE abundances for Nd II and Nd III lines in A type stars, and found much larger corrections than the Sun up to ${\sim} 1$\,dex which intensified with increasing stellar effective temperatures. They found that NLTE calculations were able to decrease the abundance difference between \ndii\ and \ndiii\ lines observed in chemically peculiar Ap stars, significantly reducing the LTE ionization imbalance between first and second ionized species.
Nevertheless, NLTE effects on Nd in FGK \rproc\ enhanced metal-poor stars have not been tested for either the optical or NIR wavelength ranges yet to our knowledge. Additionally, recent models show that NLTE effects may indeed be non-negligible for ionized dominant species of other heavy neutron-capture elements like yttrium \citep{storm2023}, europium \citep{guo2025}, strontium, and barium \citep{mashonkina2023}. 

In this paper, we thus present a novel \ndi/\ndii\ model atom which we thoroughly test for NLTE abundance analysis in FGK type stars, over a wide range of stellar parameters and \AB{Nd}{Fe} abundance ratios. Section \ref{sec:modelatom} details the construction and calibration of this model atom, and in Section \ref{sec:nlte_analysis}, we use our Nd model atom to analyze NLTE abundances in a sample of well-studied metal-poor stars and demonstrate the necessity of calculating NLTE corrections for a wide range of stellar parameters. Following this, we analyze a grid of such corrections for metal-poor FGK star parameters and discuss the physical implications of these calculations in Section \ref{sec:results}.

\section{The Model Atom}\label{sec:modelatom}

\subsection{Construction from Atomic Data}\label{sec:data_collection}

To calculate NLTE Nd abundances in stars, a model atom must first be assembled to account for as many transitions as possible, both radiatively and collisionally induced. Energy levels, electron configurations, and transition probabilities for this model were collected from existing literature and experimental atomic databases. To generate photoionization cross-sections and collisional information for the model atom, we use the Python based code \texttt{FORMATO3} \citep{merle2011}\footnote{\url{https://github.com/thibaultmerle/formato3}}. \texttt{FORMATO3} is an efficient and fast code that is useful for generating large model atoms that are compatible with the NLTE radiative transfer code \texttt{MULTI2.3} \citep{Carlsson1986,carlsson1992} which we used in this work.

\begin{figure*}[ht!]
\begin{center}
\vspace*{0cm}
\includegraphics[scale=0.57]{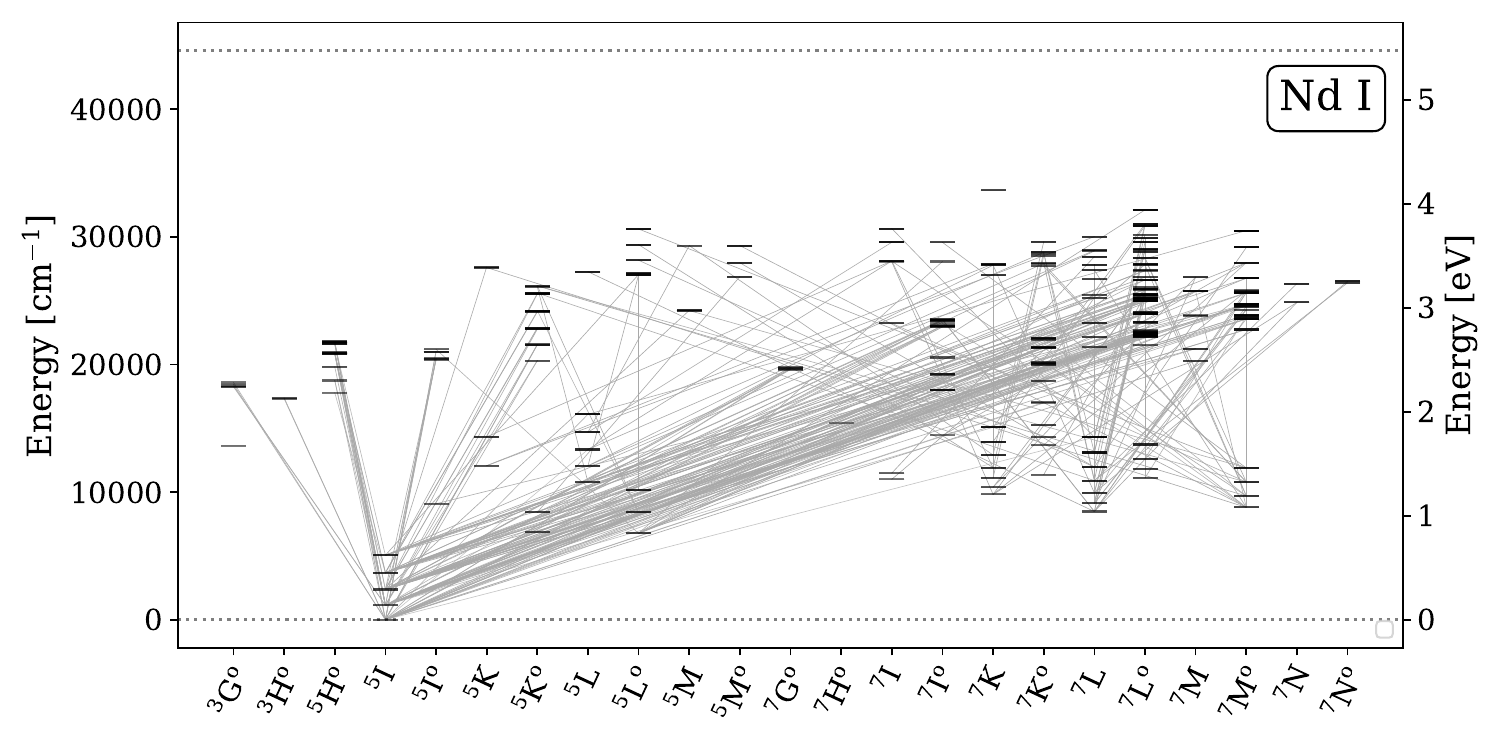}
\includegraphics[scale=0.57]{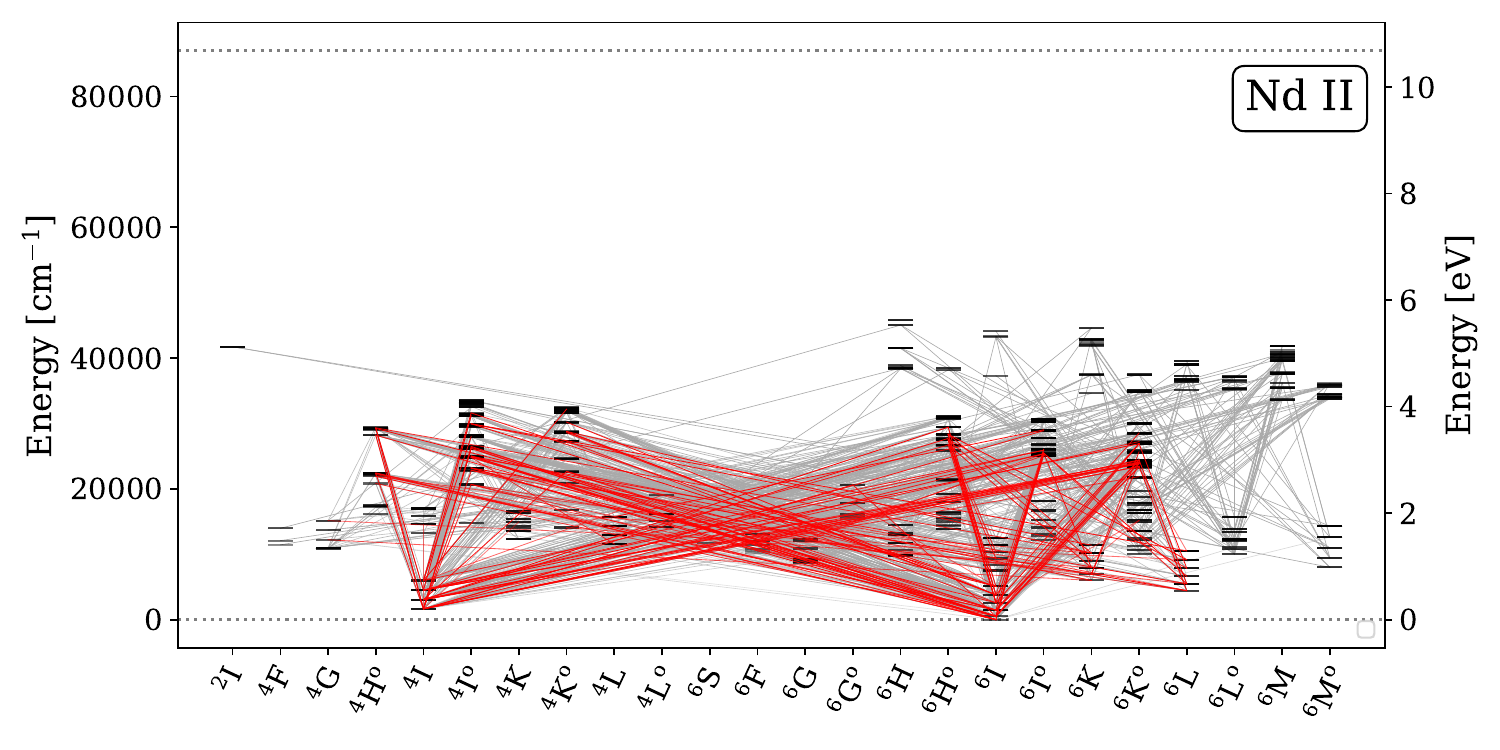}
\caption{\label{fig:grotrian}Grotrian diagrams of \ndi{} and \ndii{} displaying all energy levels and bound-bound transitions as generated by \texttt{FORMATO3}. Spectral line data were assembled from VALD, NIST, linemake, and \citet{Hasselquist_2016}. Dotted lines indicate the ground state and ionization energy of each species, and 122 transitions of interest that have been previously observed in stellar spectra are highlighted in red.}
\end{center}
\end{figure*}

Spectral lines for \ndi{} and \ndii{} were extracted from the Vienna Atomic Line Database (VALD) \citep[and references therein]{DREAM1, VALD_Nd_1, VALD_Nd_2, vald4}\footnote{\url{http://vald.astro.uu.se/~vald/php/vald.php}}. Energy levels from VALD were complemented by cross-referencing with \ndi{} and \ndii{} energy levels obtained from the NIST Atomic Spectral Database \citep{NIST_NdI_levels, Jean_Blaise_1984, Johnson2017}\footnote{\url{https://physics.nist.gov/PhysRefData/ASD/levels_form.html}}. We also implemented reliable $\log(gf)$ values from the open source database linemake \citep{linemake, denhartog2003, roederer_2008}\footnote{\url{https://github.com/vmplacco/linemake}} whenever possible. Details of the energy levels and bound-bound transitions are presented in the Grotrian diagrams of Figure \ref{fig:grotrian}. In this figure, 114 optical transitions and 8 NIR transitions are highlighted in red; these transitions are of particular interest for this study, as they have been previously observed in stellar spectra (sources for these observations are provided later in Table \ref{tab:star_list}). The total numbers of levels, transitions, and collisions in the model atom are given in Table  \ref{tab:model_atom}.

\begin{deluxetable}{ccccc}
\tablecaption{\label{tab:model_atom}Makeup of Nd Model Atom}

\tablehead{\colhead{} & \colhead{\ndi} & \colhead{\ndii} & \colhead{\ndiii} & \colhead{Total}} 

\startdata
Levels & 268 & 556 & 1 & 825 \\
BB Transitions & 281 & 1277 & {} & 1558 \\
BF Transitions & 268 & 556 & {} & 824 \\
Electron BB Collisions & 35778 & 154290 & {} & 190068 \\
Electron BF Collisions & 268 & 556 & {} & 824 \\
Hydrogen BB Collisions & 281 & 1277 & {} & 1558 \\
Hydrogen BF Collisions & 268 & 556 & {} & 824 \\
\enddata

\tablerefs{\citet{merle2011}, \citet{Hasselquist_2016}, VALD}
\end{deluxetable}

Our final constructed model atom contains 268 \ndi\ levels and 556 \ndii\ levels, and is closed by the ground state of \ndiii. Additionally, a total of 1550 bound-bound transition lines were included from VALD, including 281 \ndi\ lines and 1269 \ndii\ lines. We also added 8 NIR \ndii\ lines from \citet{Hasselquist_2016} for analysis purposes.

For bound-free transitions, we used the classical approximation following Kramer's law \citep{kramer1968} to approximate the photoionization cross-section as a function of frequency:

\begin{displaymath}
    \sigma^{ic}_\nu = \sigma^{ic}_{\nu0}\left(\frac{\nu}{\nu_0}\right)^{-3}
\end{displaymath}

where $\nu$ is the photon frequency, and $\nu_0$ is the minimum threshold frequency required to ionize the energy level $i$, whose photoionization cross section $\sigma^{ic}_{\nu0}$ (in $\mathrm{cm}^{-2}$) is given by:

\begin{displaymath}
    \sigma^{ic}_{\nu0} = 7.91 \times 10^{-22} \frac{n}{Z^2}g^{ic}_{\text{rad}}
\end{displaymath}

where $n$ is the principal quantum number of the electron configuration of the lower level $i$, $Z$ is the atomic number of the element ($Z=60$ for Nd), and $g^{ic}_{\text{rad}}$ is the Gaunt factor, a corrective factor accounting for departure from classical physical models \citep{Ezzeddine2016a} which \texttt{FORMATO3} calculates using results from \citet{menzel1935}, \citet{karzas1961}, and \citet{janicki1990}.

We also coupled all of our energy levels via inelastic electron and hydrogen collisional transitions. For electron collisional cross-sections, the Seaton approximation was used to calculate both allowed transitions \citep{seaton1962a} and forbidden transitions \citep{seaton1962b}. For inelastic hydrogen collisions, we used the Drawin approximation \citep{drawin1968, drawin1969a}, multiplied by a scaling factor (\sh).

\subsection{Model atmospheres and radiative transfer codes}

In this work, our analysis utilizes the NLTE radiative transfer code \texttt{MULTI2.3} to generate curves of growth for abundance analysis and the computation of NLTE corrections detailed in Section \ref{sec:nlte_correction_grid}. \texttt{MULTI2.3} is a widely used open-source radiative transfer code used to compute NLTE and LTE level populations and NLTE line profiles. 
\texttt{MULTI2.3} uses background opacities from the 1973 Uppsala package (\citenp{gustafsson2008} and references therein) to model the spectral continuum. As we are primarily concerned with FGK stars, we adopt the widely used 1D, LTE MARCS grid of model atmospheres \citep{gustafsson2008} for analysis, with spherical models for giant stars (\logg\ $\lesssim 3.0$) and plane-parallel models for dwarf stars (\logg\ $\gtrsim 3.0$). We adopt model atmospheres with the alpha-enhanced ``standard'' mixture of metals as defined in the MARCS documentation\footnote{\url{https://marcs.astro.uu.se/docs.html}}. 
These models use the $\alpha$ element abundances \AB{$\alpha$}{Fe} $= 0.0$ for \feh\ $\ge 0.00$, \AB{$\alpha$}{Fe} $= +0.4$ for \feh\ $\le -1.00$, and \AB{$\alpha$}{Fe} scaling linearly between these two values for $-1.00 \le$ \feh\ $\le 0.00$.
We interpolated MARCS model atmospheres for adopted stellar parameters (\teff, \logg, \feh, \vt) using the routine \texttt{interpol\_marcs.f}, created by Thomas Masseron\footnote{\url{https://marcs.astro.uu.se/software.html}}.

To generate the NLTE synthetic spectra, we used the NLTE-enabled spectral synthesis code \texttt{Turbospectrum} (TS-NLTE; \citenp{plez2012,gerber2023}) as explained in Section \ref{sec:synthetic}. The code was updated with a NLTE correction grid centered around the star's parameters, generated by \texttt{MULTI2.3} using our new \ndi/\ndii\ model atom. We note that while we use TS-NLTE for comparison between synthetic and observed spectra as shown in Figure\,\ref{fig:synthetic_spectra}, we do not use it to fit or derive abundances.

\subsection{Testing on the Sun \& Calibrating \sh} \label{sec:testing_sun}

With the Nd model atom generated as described above, we proceeded by deriving NLTE and LTE Nd abundances for the Sun.
We determine the abundances of each Nd II line using a curve of growth (COG) method based on the computed EW for each input \logeps\ value. 
\texttt{MULTI2.3} computes departure coefficients, allowing for comparison between the NLTE and LTE abundances for each line. 

\begin{figure}[ht!]
\begin{center}
\vspace*{0cm}
\includegraphics[scale=0.64]{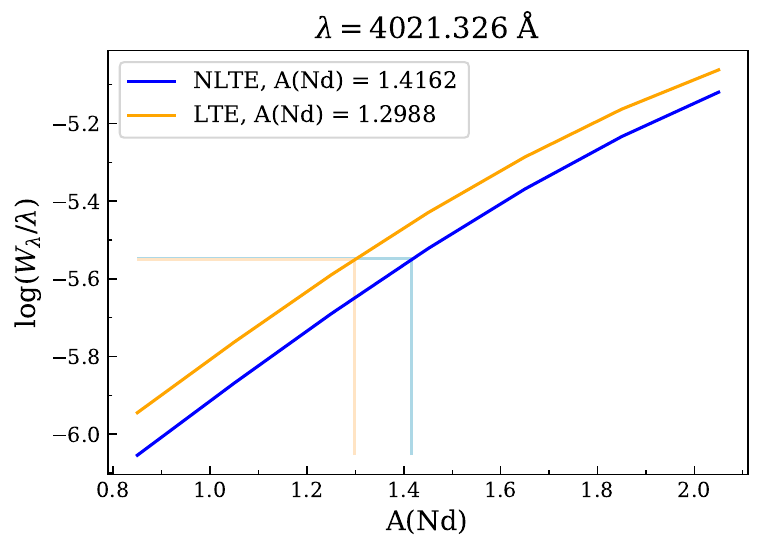}
\includegraphics[scale=0.64]{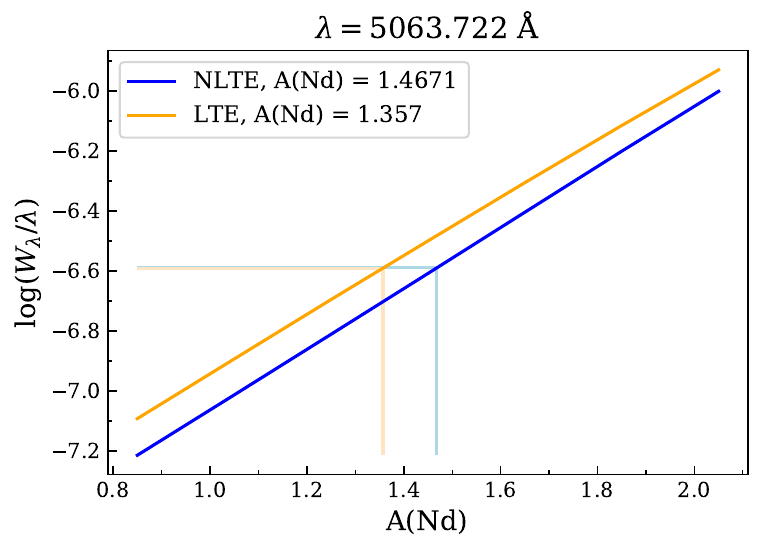}
\caption{\label{fig:growth_curves}Curves of growth for two \ndii\ lines generated with \texttt{MULTI2.3} using the solar MARCS model. Measured EWs from \citet{denhartog2003} were converted into \logeps\ values in both NLTE (dark blue) and LTE (light orange) by interpolating from a cubic best-fit curve.}
\end{center}
\end{figure}

\begin{figure}[t!]
\begin{center}
\vspace*{0cm}
\hspace*{-.2cm}
\includegraphics[scale=0.53]{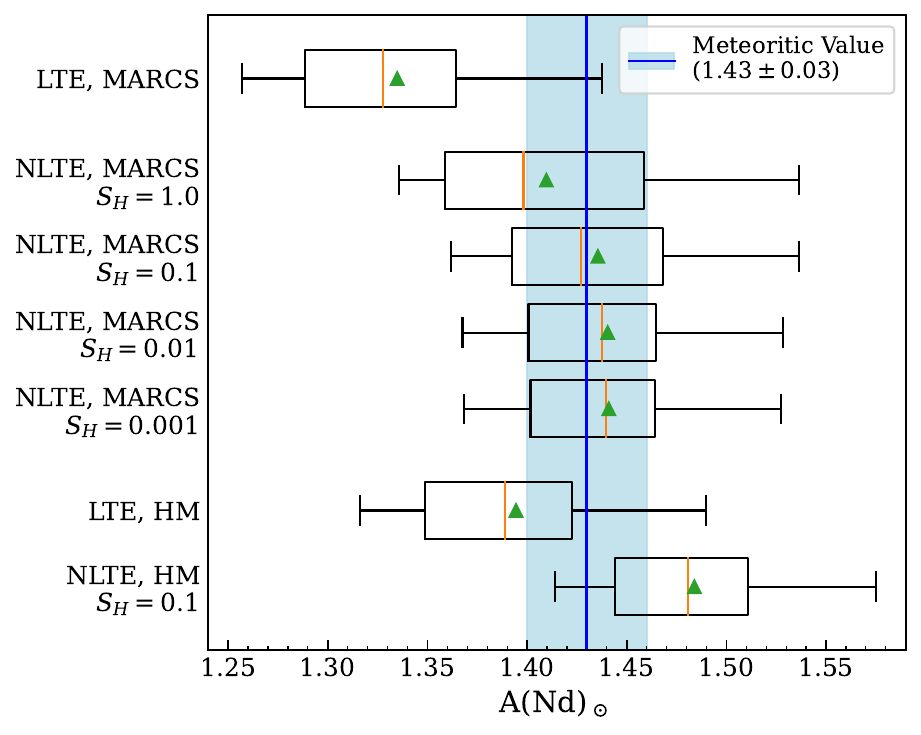}
\includegraphics[scale=0.53]{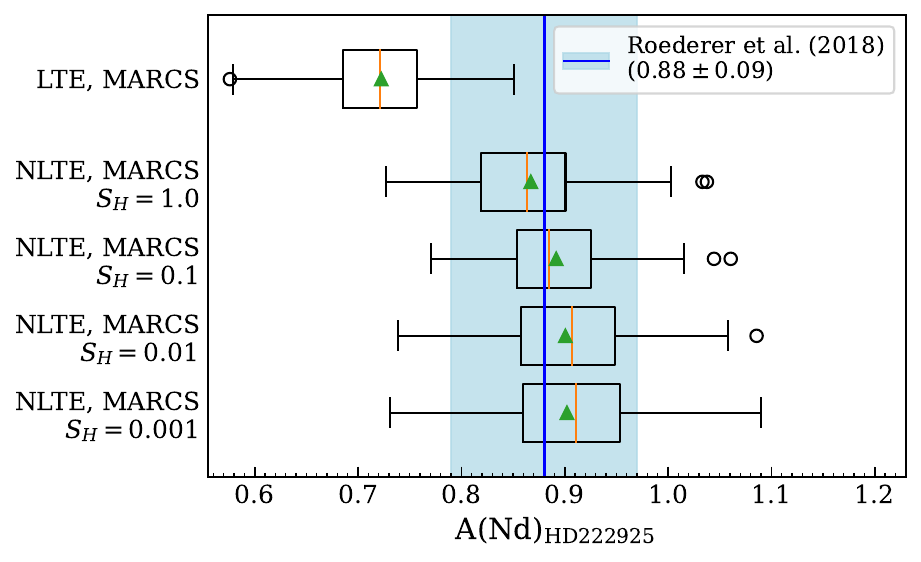}
\caption{\label{fig:boxplots}{\ndii\ abundance calculations based on observed lines for the Sun (top) and HD222925 (bottom) for LTE and NLTE models with different \sh\ factors, compared to published A(Nd) values (shown as blue shaded regions). Vertical axis labels indicate whether the HM model or a MARCS model atmosphere were used for different \sh\ factors. Orange lines show the median of the line-by-line Nd abundances, while green triangles show the mean. The lower and upper boundaries of the boxes represent the first and third quartiles of the line-by-line abundances respectively, with outliers shown as black circles.}}
\end{center}
\end{figure}

For the solar abundance analysis, we chose 25 optical \ndii\ lines (listed in Appendix Table~\ref{tab:ews_list}) with EWs measured by \citet{denhartog2003} from a solar spectrum collected with the 1-meter Fourier Transform Spectrometer at the National Solar Observatory. 
Figure~\ref{fig:growth_curves} shows two example COGs computed for the two solar \ndii\ lines at 4021\,{\AA} and 5063\,{\AA}, respectively. 
The figure also shows the observed EW for each line and the abundance it corresponds to for both COGs; the horizontal distance between the two COGs gives the NLTE correction for a particular EW. The final solar Nd abundance was calculated by averaging the abundances derived from all 25 solar lines, to determine \logeps$_{\odot}$. Figure \ref{fig:boxplots} shows average solar abundances derived for LTE and NLTE models employing different $S_H$ factors ranging from 0.001 to 1.0, using both MARCS and the solar Holweger \& Mueller model (HM; \citenp{HM1974}). In addition, we generated similar boxplots from 98 observed \ndii\ lines for HD222925 -- a highly enhanced \rproc\ benchmark star with a previously derived LTE Nd abundance from \citet{roederer2018}. For both stars, our derived abundances are higher with the NLTE models as compared to LTE.
This is in line with predictions of NLTE abundances for dominant species in FGK type atmospheres \citep{bergemann2012,lind2012,ezzeddine2017}.

For the Sun, we derive an LTE Nd abundance of $1.335 \pm 0.053$ with a MARCS model atmosphere. Our derived NLTE values of \logeps$_{\odot}$ with MARCS models are $1.410 \pm 0.056$, $1.436 \pm 0.050$, $1.440 \pm 0.048$, and $1.441 \pm 0.047$ for \sh\ =  1.0, 0.1, 0.01, and 0.001, respectively. These are in close agreement with the meteoritic value of $1.43 \pm 0.03$ \citep{grevesse2007}. For HD222925, we derive an LTE Nd abundance of $0.722 \pm 0.057$. Our derived NLTE abundances are $0.867 \pm 0.061$, $0.892 \pm 0.058$, $0.901 \pm 0.070$, and $0.902 \pm 0.073$ for \sh\ =  1.0, 0.1, 0.01, and 0.001, respectively. We provide the LTE abundance from \citet{roederer2018} to show the size of these error bars, but note that our values were computed using a different MARCS model atmosphere with re-derived NLTE stellar parameters that are discussed further in Section \ref{sec:metal_poor_star}. As both stars yield a small spread in average A(Nd) as a function of \sh, any abundance uncertainty arising from the choice of \sh\ is likely negligible compared to other sources of uncertainty. A full list of the line-by-line solar Nd abundances determined for each \sh\ value, as well as the average abundances, are given in the Appendix Table~\ref{tab:ews_list}. 

It is worth noting that our derived solar LTE abundance differs from the abundance reported in \citet{denhartog2003} by ${\sim}0.1$\,dex. This difference likely arises from the different model atmosphere and radiative transfer codes used in both studies; \citet{denhartog2003} used the empirical 1D Holweger \& Mueller model atmosphere \citep{HM1974}, while our analysis employs a theoretical MARCS solar 1D, LTE model. For the 25 common solar lines analyzed in this work, the mean and standard deviation of the reported abundances in \citet{denhartog2003} is \logeps\ $= 1.44 \pm 0.04$. For comparison, we ran \texttt{MULTI2.3} with the Holweger \& Mueller model atmosphere, for which we derived \logeps\ $= 1.40 \pm 0.05$ in LTE and \logeps\ $= 1.48 \pm 0.05$ in NLTE for the same 25 lines. Both of these abundances agree with the value from \citet{denhartog2003} to within $1\sigma$.

\subsection{Testing Atom Completeness}\label{sec:completeness}

\begin{figure*}[ht!]
\begin{center}
\vspace*{0cm}
\hspace*{0cm}
\includegraphics[scale=0.57]{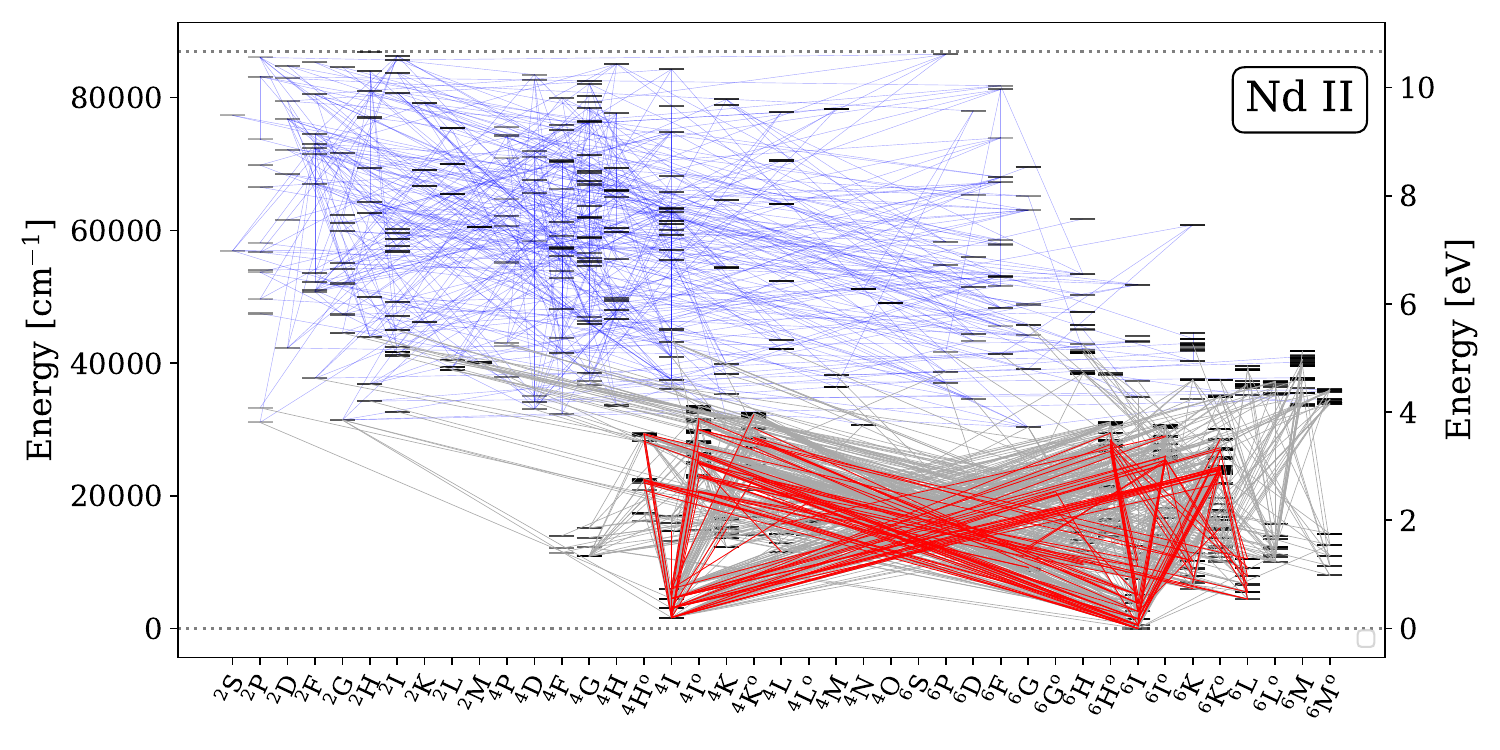}
\includegraphics[scale=0.57]{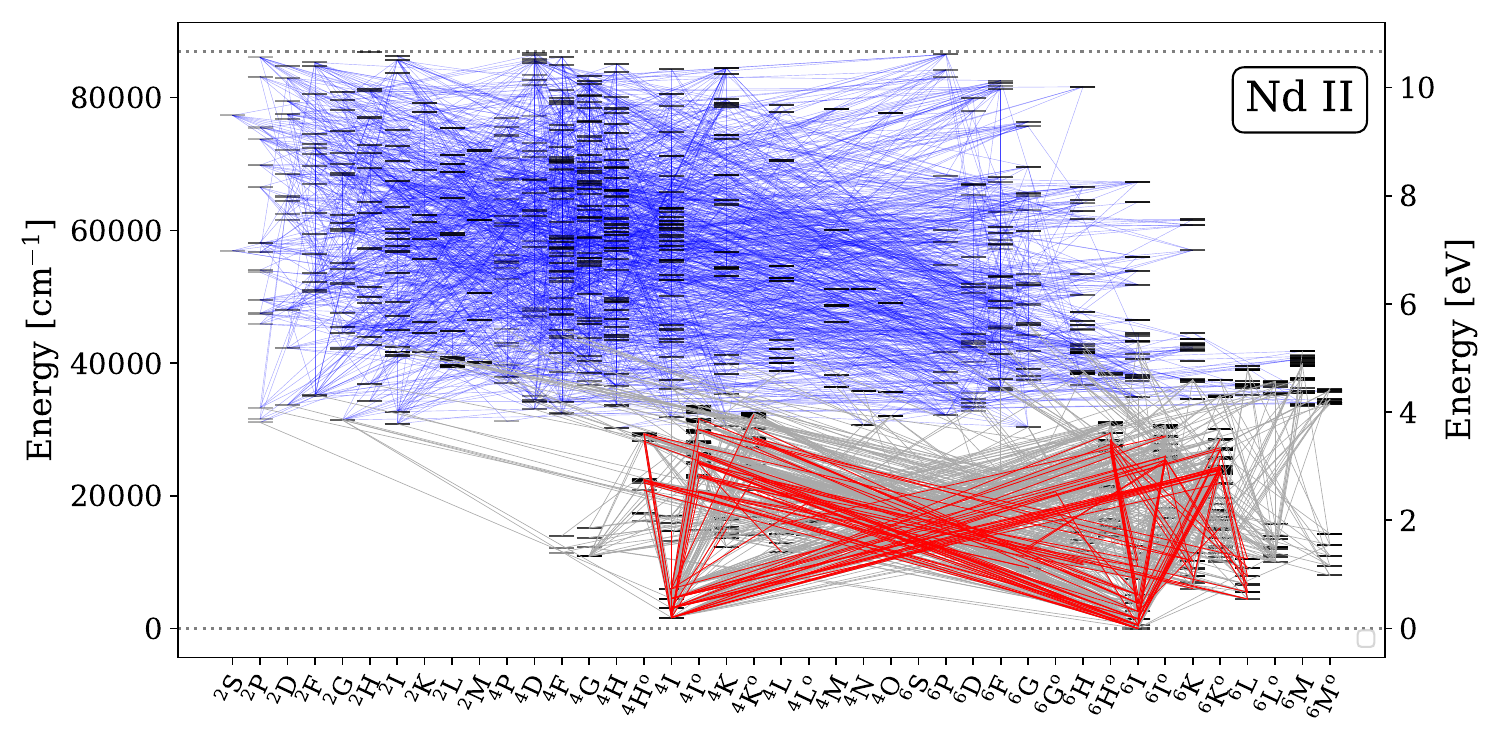}
\includegraphics[scale=0.57]{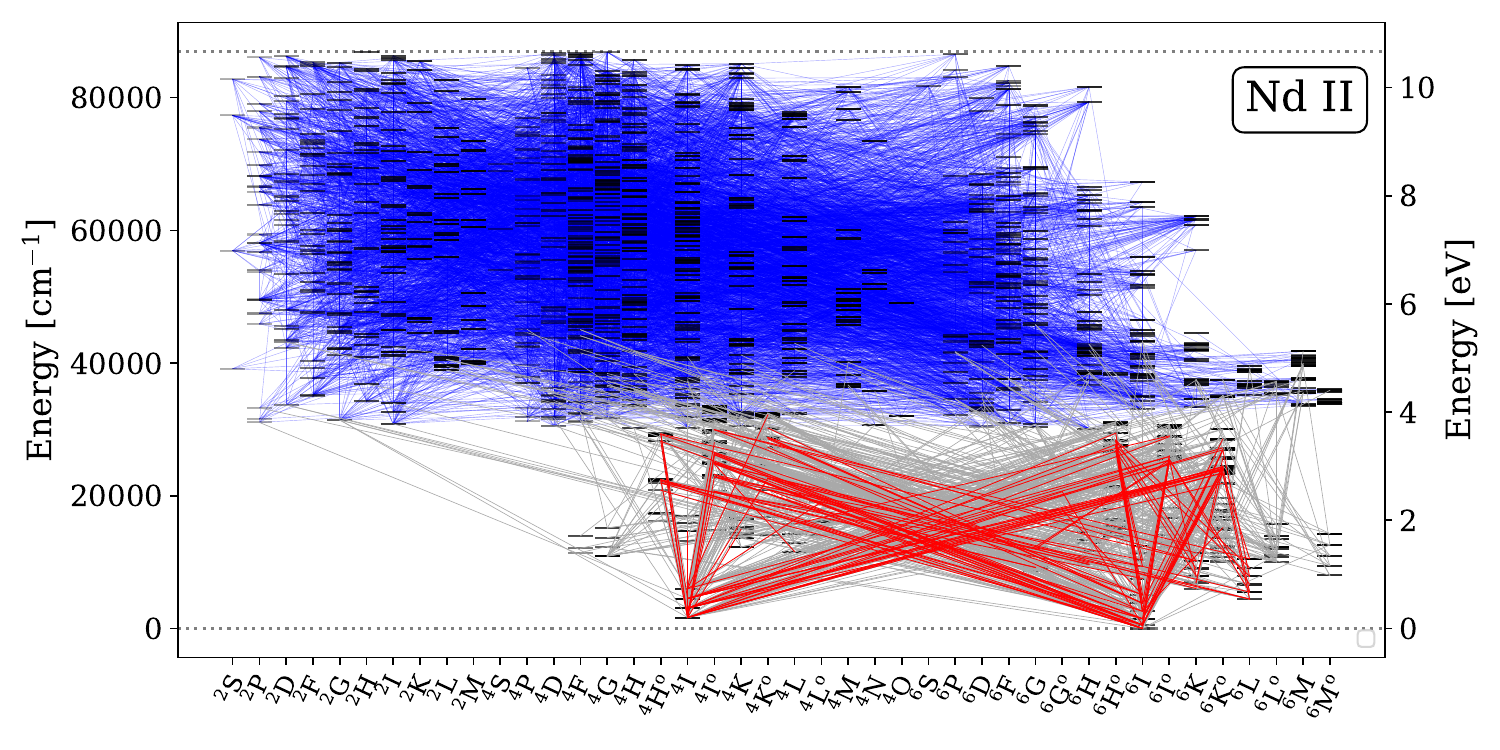}
\caption{\label{fig:grotrian_theoretical}Grotrian diagrams of \ndii\ for three model atoms: Atom A (top), Atom B (middle), and Atom C (bottom) (see Section\,\ref{sec:completeness} for details on the atoms). Dotted lines indicate the ground state and ionization energy of Nd II. The 122 transitions of interest shown in Figure \ref{fig:grotrian} are also highlighted here in red. Transitions are highlighted in blue if their lower energy level is theoretical.}
\end{center}
\end{figure*}

We note that the highest experimental \ndii\ energy level from VALD and NIST  is 45801.380 $\mathrm{cm}^{-1}$, which is significantly lower than the second ionization energy of Nd at 86970 $\mathrm{cm}^{-1}$. To test whether this large energy gap affects our NLTE results, we created three additional versions of the Nd model atom described in Section\,\ref{sec:data_collection}, henceforth referred to as atoms A, B, and C. These models supplement the 556 \ndi\ energy levels and 1277 \ndii\ lines from VALD and NIST with theoretically computed \ndii\ levels and lines from \citet{Gaigalas2019}. For atom A, we  added 251 theoretical levels and 162 new lines; Atom B adds 515 theoretical levels and 2009 new lines; Atom C adds 1055 theoretical levels and 7967 new lines. Grotrian diagrams of \ndii\ for these three model atoms are shown in Figure \ref{fig:grotrian_theoretical}. The added theoretical levels have energies above 30000 $\mathrm{cm}^{-1}$, and all three subsets of levels were chosen to roughly preserve the relative density of levels as a function of energy. To construct Atom A, we first arranged all theoretical levels above 30000 $\mathrm{cm}^{-1}$ by energy and sampled 251 levels at evenly spaced indices. We then compiled a list of every theoretical bound-bound transition involving only these 251 levels, undersampling theoretical transitions as necessary in order to make computations tractable and avoid overloading \texttt{MULTI2.3}. We reviewed these transitions to ensure sufficient coupling between experimental and theoretical levels, and then we added the theoretical levels and transitions to our compiled experimental data to create the final Atom A. This same procedure was used to construct Atoms B and C, with the only difference being the number of energy levels sampled, as shown in Table \ref{tab:testing_completeness}.

\begin{deluxetable}{ccc}
\tablecaption{\label{tab:testing_completeness}Mean NLTE corrections for the Sun and HD222925 calculated from model atoms supplemented with theoretical data.}
\tablehead{\colhead{} & \colhead{\dnlte} & \colhead{\dnlte} \\
{} & (Sun) & (HD222925)} 
\startdata
Experimental Data Only & $0.101 \pm 0.017$ & $0.168 \pm 0.052$ \\
Atom A (+251 levels) & $0.099 \pm 0.017$ & $0.165 \pm 0.050$ \\
Atom B (+515 levels) & $0.099 \pm 0.017$ & $0.165 \pm 0.087$ \\
Atom C (+1055 levels) & $0.098 \pm 0.016$ & $0.165 \pm 0.088$   \\
\enddata
\end{deluxetable}

For each of these three model atoms, we re-derived NLTE abundances for the 25 solar \ndii\ lines mentioned above, as well as for 98 \ndii\ lines from HD222925 -- a highly enhanced \rproc\ benchmark star with a previously derived LTE Nd abundance from \citet{roederer2018}. Table \ref{tab:testing_completeness} shows the average and standard deviation of the derived \logepsNLTE\ $-$ \logepsLTE\ value (i.e. the NLTE correction, henceforth referred to as \dnlte) from spectral lines in both stars. The differences in NLTE corrections between different model atoms are negligible and the spread of line-by-line NLTE corrections stays within a reasonable margin of error for abundance analysis, indicating that NLTE modeling effects do not change significantly with the addition of higher energy theoretical levels and associated lines. Even in the case that the current data from VALD only accounts for ${\sim}1/3$ of all \ndii\ levels and ${\sim}1/6$ of all \ndii\ lines (Atom C), the NLTE modeling effects for currently known lines in both the Sun and \rproc\ enhanced star HD222925 are unaffected. Thus, going forward, we adopt the model atom constructed from experimental VALD and NIST data as our final atom for the rest of this study.

As another test of our NLTE model, we computed NLTE corrections for a small set of model atmospheres with typical A-type main-sequence parameters (\logg\ $= 4.0$, \feh\ $= -1.50$, \vt\ $= 1.0$ \kms, $7500\ \mathrm{K} \le$ \teff\ $\le 8000\ \mathrm{K}$) with an Nd enhancement of \AB{Nd}{H} $ = 2.5$. These stellar parameters are outside the scope of our work, but the goal of this analysis was to compare \dnlte\ values from our model atom to those from \citet{mashonkina2005}, which were calculated with theoretical atomic data. For the \ndii\ lines at $\lambda = 4061$, $4706$, and $5319$ \AA, our corrections are in agreement with \citet{mashonkina2005} to within a typical margin of error for abundance analysis (${\pm}0.2$\,dex) and trend in the same direction as \teff\ changes. As expected, we find relatively small \dnlte\ values, increasing from roughly $-0.1$ to $0.1$ dex as \teff\ increases from 7500 K to 8000 K.

Additionally, we added hyperfine structure (HFS) and isotopic splitting data for six \ndii\ lines from \citet{roederer_2008} to our model atom to investigate the impacts of HFS and isotopic splitting on NLTE abundance analysis, which have previously been assumed to be negligible \citep{denhartog2003}. Re-deriving NLTE abundances for the Sun and HD222925 with this model atom confirms this hypothesis; for both stars, line-by-line Nd abundances changed by less than 0.005\,dex, and so we proceed without further implementation of HFS data and isotopic splitting.


\section{NLTE Abundance Analysis}\label{sec:nlte_analysis}

\subsection{NLTE Synthetic Spectra}\label{sec:synthetic}

We generated synthetic spectra from our model atom using TS-NLTE in order to compare with existing observations of stars.
We used the Python wrapper \texttt{TSFitPy} to run TS-NLTE in LTE and NLTE \citep{gerber2023, storm2023} with 1D MARCS model atmospheres. We implemented isotopic ratios of neutron-capture elements up to europium ($Z = 63$) from \citet{sneden2008} in \texttt{TSFitPy}. 
Our atomic line lists for synthesizing and fitting the spectral regions around \ndii\ lines of interest come from Gaia-ESO \citep{heiter2021}. 

\begin{figure*}[ht!]
\begin{center}
\includegraphics[scale=0.55]{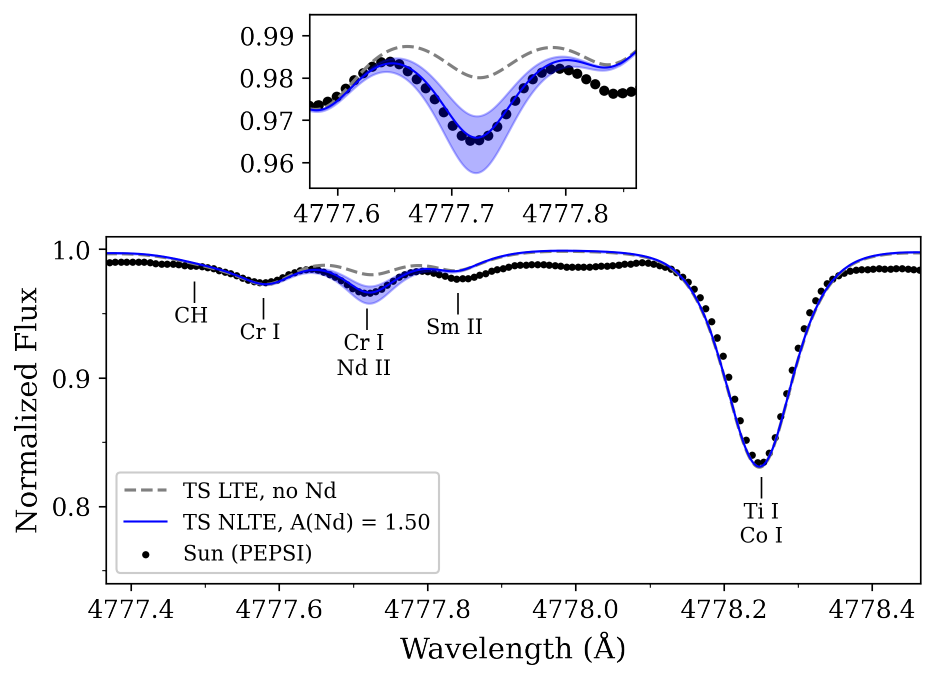}
\hspace{0.3cm}
\includegraphics[scale=0.55]{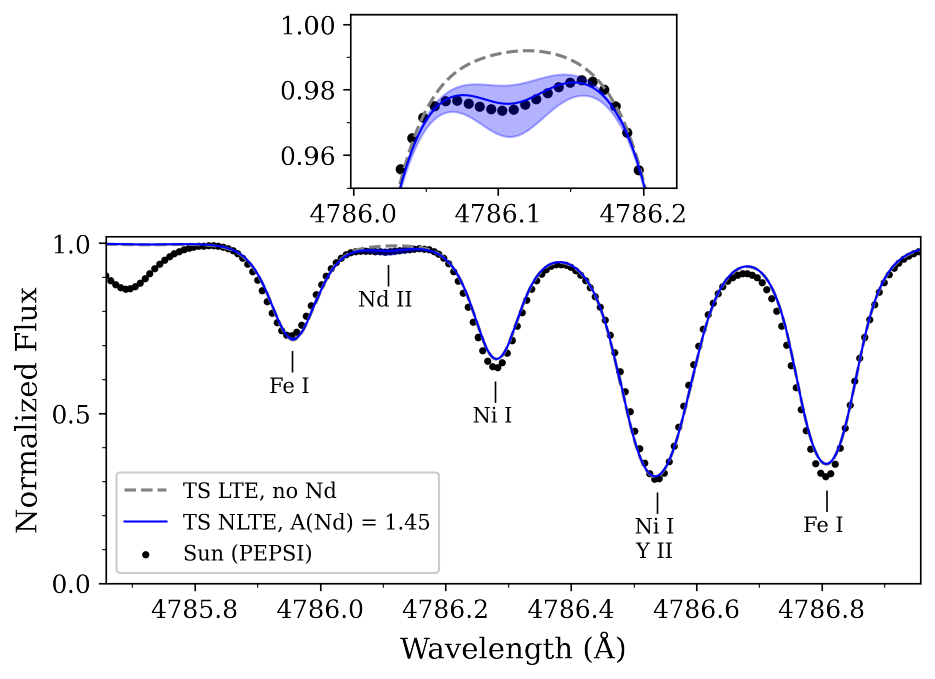} \\
\vspace*{0.5cm}
\includegraphics[scale=0.55]{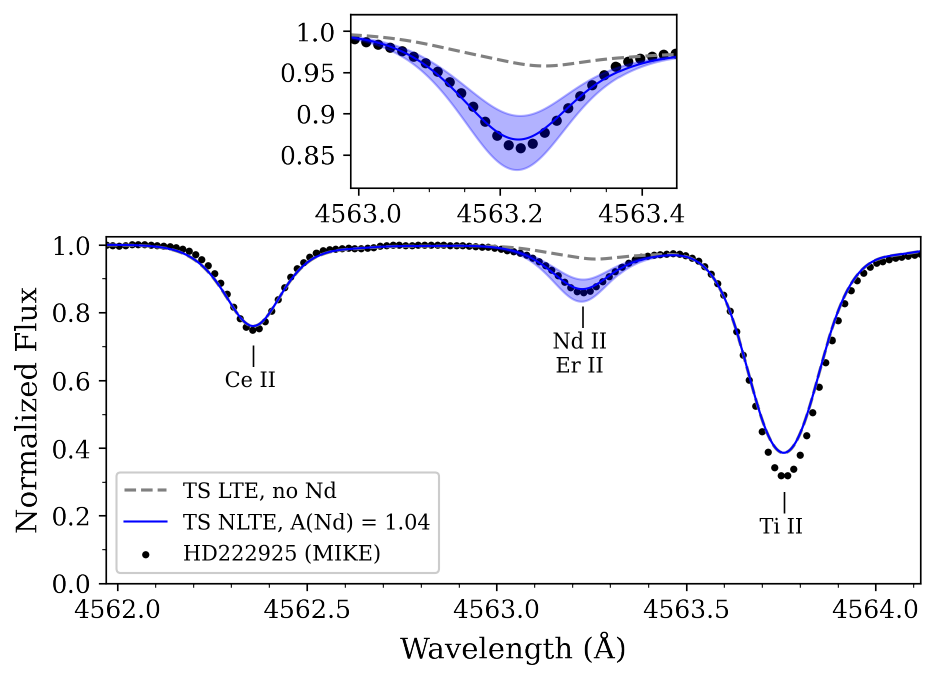}
\hspace{0.3cm}
\includegraphics[scale=0.55]{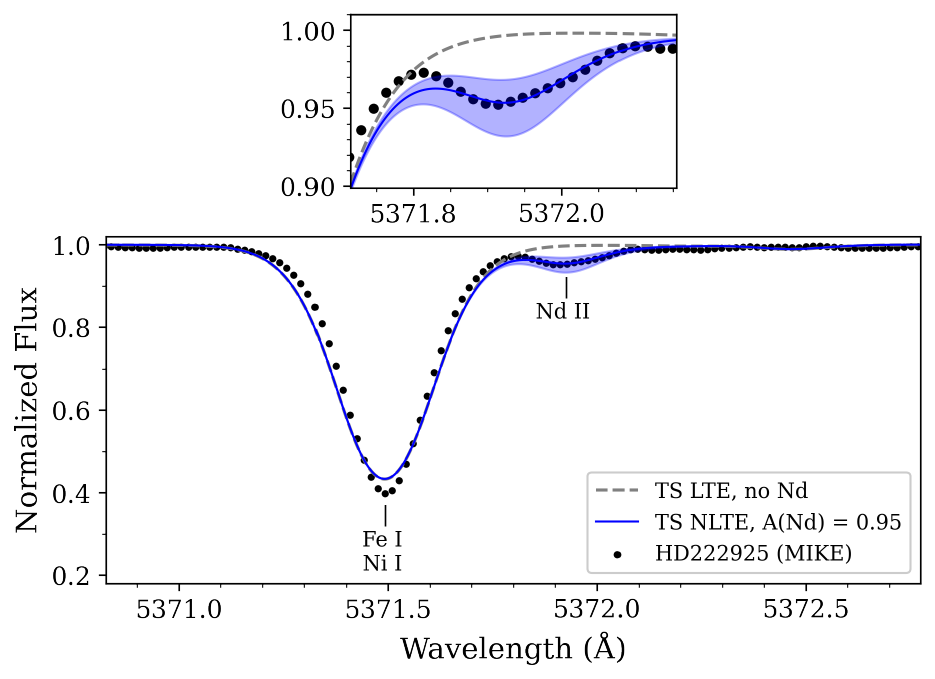}
\caption{\label{fig:synthetic_spectra} Observed and synthetic spectra for wavelength regions around two solar \ndii\ lines (top) and two \ndii\ lines in HD222925 (bottom). For HD222925, abundances were obtained and synthetic spectra were generated using the stellar parameters listed in \citet{roederer2018}. Shaded regions indicate $\pm0.2$\,dex in the NLTE abundance.
Prominent spectral lines and features are labeled. 
Small panels above each plot are zoomed in on the \ndii\ lines under consideration.}
\end{center}
\end{figure*}

For the solar synthetic spectra, we use \vt\ $= 0.85$ \kms\ and a macroturbulent velocity of 3 \kms \citep{doyle2014}. For the synthetic spectra of HD222925, we adopt \vt\ $= 2.2$ \kms\ from \citet{roederer2018}, and a macroturbulent velocity of 7 \kms\ (I. U. Roederer, private communication). Figure \ref{fig:synthetic_spectra} shows the generated synthetic spectra plotted over selected  \ndii\ lines observed in the Sun and HD222925. Details on the spectrum and stellar parameters for HD222925 can be found in \citet{roederer2018}, and the solar spectrum was taken in 2016 with the Potsdam Echelle Polarimetric and Spectroscopic Instrument (PEPSI; \citenp{strassmeier2018}). 
In this figure,  we show wavelength ranges around \ndii\ spectral lines for both the Sun and HD222925. To ensure the best fit to the synthetic spectra, we renormalize the observed spectra within each range in blended regions. The synthetic spectrum shown in each plot uses NLTE modeling for \ndii\ with the respective abundances we derived using \texttt{MULTI2.3}, and shaded regions showing a typical abundance uncertainty of 0.2\,dex. LTE is assumed for all other elements; solar abundances are from \citet{bergemann2021} and \citet{magg2022}, and mean abundances for HD222925 are adopted from \citet{roederer2018}. 
Some of these lines are blended, but a synthetic spectrum with no Nd (shown as a dashed line) is also included to demonstrate the contribution of Nd to each spectral feature. By visual inspection, it can be seen that the \ndii\ lines in our synthetic spectra agree fairly well with the observed spectrum. Thus, we are confident that our model atom is capable of reproducing synthetic spectra (using previously published stellar parameters) that agree with observational data, and so we proceed with more detailed NLTE analysis.

\subsection{Application to Metal-Poor Stars}\label{sec:metal_poor_star}

After testing our NLTE model atom for the Sun and a metal-poor \rproc\ enhanced benchmark star (HD222925), we proceed to use our model to derive NLTE abundances for a sample of metal-poor stars with previously measured \ndii\ LTE abunbdances from the literature. The list of stars chosen are given in Table \ref{tab:star_list} along with their stellar parameters from referenced literature sources. Uncertainties on stellar parameters are also taken from the papers in the table; if no uncertainty is provided in the paper, we instead adopt the typical value provided by the paper for each stellar parameter.
LTE and NLTE \ndii\ abundances were derived for each star from observed EWs collected from the literature as listed in Table \ref{tab:star_list}.
Most of these studies list a typical EW uncertainty on the order of ${\sim}$1 m\AA, but do not report a measurement uncertainty for every individual line. For the Sun and HD222925, we performed robustness tests, increasing and decreasing each reported EW by a conservative uncertainty estimate of 10\% to observe how the derived abundance changes. We find that this produces an abundance uncertainty of ${\sim}$0.05\,dex. As this is lower than typical stellar abundance uncertainties, we expect that other factors such as the choice of model and stellar parameters likely have a larger effect on the overall line-by-line uncertainty for our sample.

Abundances were derived using stellar parameters from the references listed in Table \ref{tab:star_list}, and MARCS model atmospheres corresponding to the parameters of each star. For each star, we adopt typical uncertainties from the same paper as stellar parameters, if no uncertainties are reported. For stars with 6 or more measured EWs, we provide the standard deviation of the line-by-line abundances as the uncertainty on \logeps; for stars with fewer than 6 measurements, we instead adopt a typical uncertainty of $\pm 0.2$\,dex (for HD122563, we adopt $\pm 0.1$\,dex due to the higher signal-to-noise ratio of the observations).
For most stars, EW measurements were available and used to derive line-by-line \logeps\ abundances using COGs as described in Section \ref{sec:testing_sun}. However, whenever observed EWs were not available, we instead used the COG to synthetically derive EWs at the corresponding LTE abundances, which were then used to derive NLTE abundances for these lines. 
We note that these values do not necessarily represent real EWs measured directly from spectra, which limits the accuracy of corrections for these stars. The purpose of this method is to obtain preliminary \dnlte\ values for a few stars of interest to better probe the necessity of further NLTE analysis.
For all EWs used, the reduced equivalent width (REW, $\log(\mathrm{EW}/\lambda)$) falls on the roughly linear portion of the COG, which is most reliable for this method of analysis.

\begin{deluxetable*}{ccrrccccc}
\tablecaption{\label{tab:star_list}Benchmark Stars for NLTE Abundance Analysis of \ndii\ lines}
\tablehead{\colhead{Star} & \colhead{N} & \logepsLTE\ & \logepsNLTE\ &
\colhead{\teff\ (K)} & \colhead{\logg} & \colhead{\feh} & \colhead{\vt\ (\kms)} & \colhead{Ref.}} 
\startdata 
Sun             & 25 & $ 1.34 \pm 0.05$  & $ 1.44 \pm 0.05$  & $5777    $  & $4.44      $  & $ 0.00     $  & $0.85     $  & 1         \\
HD115444        & 34 & $-1.16 \pm 0.08$  & $-0.91 \pm 0.07$  & $4650 \pm 150$  & $1.50 \pm 0.50$  & $-2.99 \pm 0.50$  & $2.10 \pm 0.30$  & 1, 2$^a$  \\
BD+17°3248      & 55 & $-0.10 \pm 0.07$  & $ 0.11 \pm 0.07$  & $5200 \pm 150$  & $1.80 \pm 0.30$  & $-2.00 \pm 0.20$  & $1.90 \pm 0.20$  & 1, 3$^a$  \\
CS22892--052    & 35 & $-0.42 \pm 0.06$  & $-0.14 \pm 0.08$  & $4828 \pm 150$  & $1.35 \pm 0.30$  & $-3.08 \pm 0.11$  & $2.15 \pm 0.30$  & 1, 4$^a$  \\
HD122563        & 1  & $-1.88 \pm 0.10$  & $-1.59 \pm 0.10$  & $4620 \pm  92$  & $1.21 \pm 0.23$  & $-2.76 \pm 0.07$  & $1.89 \pm 0.08$  & 5, 6$^a$  \\
CS30306--132    & 6  & $-0.36 \pm 0.07$  & $-0.18 \pm 0.06$  & $5100 \pm 100$  & $2.20 \pm 0.30$  & $-2.50 \pm 0.20$  & $1.90 \pm 0.30$  & 5         \\
J0326+0202      & 1  & $-1.43 \pm 0.20$  & $-1.23 \pm 0.20$  & $5080 \pm 100$  & $2.03 \pm 0.30$  & $-3.11 \pm 0.20$  & $2.01 \pm 0.30$  & 7         \\
J1108+2530      & 2  & $-0.93 \pm 0.20$  & $-0.79 \pm 0.20$  & $5003 \pm 100$  & $2.05 \pm 0.30$  & $-2.72 \pm 0.20$  & $1.08 \pm 0.30$  & 7         \\
J1256+3440      & 2  & $-1.10 \pm 0.20$  & $-0.88 \pm 0.20$  & $5215 \pm 100$  & $1.74 \pm 0.30$  & $-2.76 \pm 0.20$  & $2.37 \pm 0.30$  & 7         \\
HD222925        & 98 & $ 0.72 \pm 0.06$  & $ 0.89 \pm 0.06$  & $5505 \pm  80$  & $2.39 \pm 0.14$  & $-1.55 \pm 0.09$  & $2.34 \pm 0.09$  & 8$^{b,c}$ \\
HD49368         & 3  & $ 2.20 \pm 0.20$  & $ 2.18 \pm 0.20$  & $3679 \pm  70$  & $0.35 \pm 0.08$  & $-0.19 \pm 0.04$  & $1.66 \pm 0.25$  & 9$^c$     \\
HD35155         & 3  & $ 3.42 \pm 0.20$  & $ 3.17 \pm 0.20$  & $3656 \pm  70$  & $0.24 \pm 0.08$  & $-0.72 \pm 0.04$  & $1.66 \pm 0.25$  & 9$^c$     \\
\enddata
\tablerefs{$^1$\citet{denhartog2003}, $^2$\citet{westin2000}, $^3$\citet{cowan2002}, $^4$\citet{frebel2013}, $^5$\citet{aoki2005}, $^6$\citet{li2023}, $^7$\citet{mardini2019}, $^8$\citet{roederer2018}, $^9$\citet{Hasselquist_2016}
}
\tablenotetext{}{\scriptsize $^{a\,}$EWs adopted from the first reference, stellar parameters adopted from the second reference. \\
$^{b\,}$Parameters recalculated in NLTE using LOTUS \citep{li2023} to reduce trends in line-by-line abundances. \\
$^{c\,}$EW measurements derived from LTE abundances using a COG.}
\end{deluxetable*}

\begin{figure}[ht]
\begin{center}
\vspace*{0cm}
\includegraphics[scale=0.65]{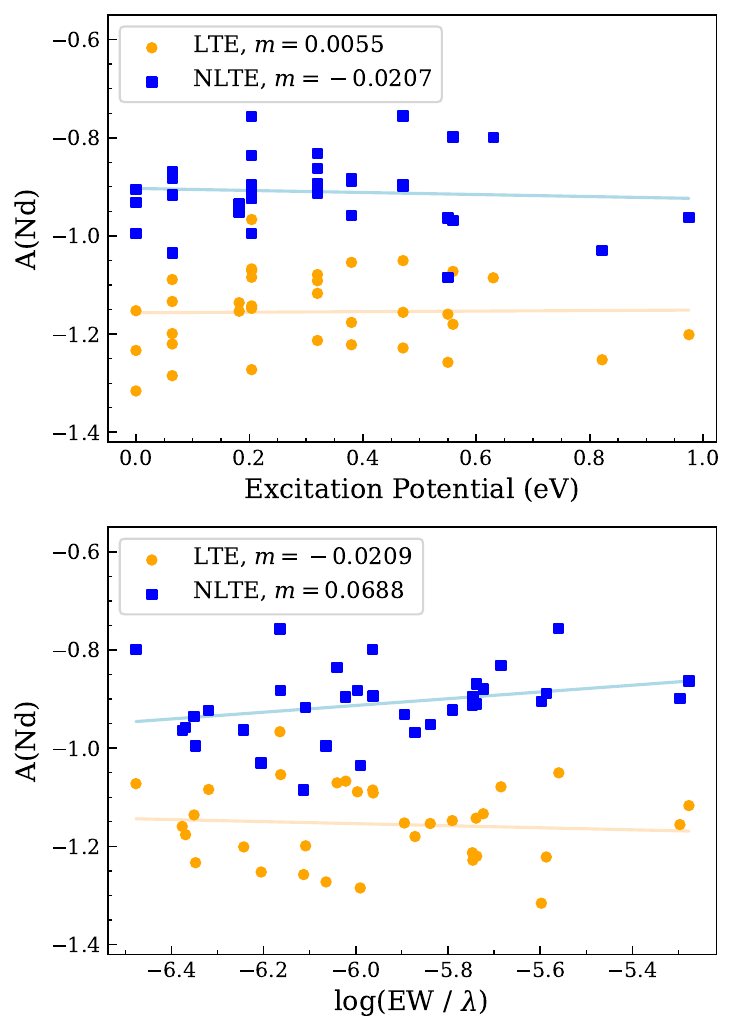}
\caption{\label{fig:diagnostics} Line-by-line \logeps\ versus excitation potential (top) and reduced equivalent widths (bottom) for HD115444, with lines of best fit for both LTE and NLTE (slopes given in legend). The LTE abundance is \logepsLTE\ $=-1.155 \pm 0.079$, and the NLTE abundance is \logepsNLTE\ $= -0.910 \pm 0.073$.}
\end{center}
\end{figure}

Upon inspecting the line-by-line \ndii\ abundance trends as a function of excitation potential (EP) and line strengths ($\log\mathrm{(EW/\lambda})$), we find that the best-fit line has a slope close to zero for most stars, for both LTE and NLTE, as demonstrated for HD115444 in Figure \ref{fig:diagnostics}. It is noteworthy to mention, however, that for HD222925 a trend was observed with a slope of $-0.079$ obtained for the NLTE line-by-line \ndii\  abundances versus EP, and of $0.068$ versus REW. To investigate whether these trends are arising due to the adopted stellar parameters, we recompute \teff, \logg, \feh\ and \vt\ using the NLTE stellar parameter optimization tool \texttt{LOTUS} \citep{li2023}, employing measured EWs for \fei\ and \feii\ lines from \citet{roederer2018}. The derived NLTE stellar parameters are \teff\ $ = 5505$, \logg\ $ = 2.39$, \feh\ $= -1.55$, and \vt\ = $2.34$ \kms, which compare relatively well to those derived by \citet{roederer2018}: \teff\ $ = 5636$, \logg\ $ = 2.54$, \feh\ $= -1.50$, and \vt\ = $2.20$ \kms. We then rederive the \ndii\ abundances using the new parameters, where we find that the overall NLTE abundance trend slopes decrease to $-0.046$ (for EP) and $0.025$ (for line strengths), respectively. Notably, these slopes are larger than those obtained for the LTE line-by-line abundances using the \citet{roederer2018} parameters, where slopes of 0.015 (versus EP) and 0.007 (versus REW) were computed.

It is possible that the slope trends in NLTE are due to  3D model atmospheric effects, which have been shown to introduce abundance trends as a function of EP for a number of elements studied in 3D, NLTE such as iron \citep{amarsi2022, li2023} and magnesium \citep{bergemann2017}. Low-EP lines are shown to be especially affected by ignoring 3D modeling effects. We are unable to diagnose these trends using a full 3D, NLTE analysis for Nd as it is beyond the scope of this current work and computationally expensive. Additionally, while some \avgNLTE\ NLTE models are publicly available for a limited number of stellar parameters \citep{gerber2023}, they are not available for stellar models of many giant stars with \teff\ $\ge 5000\ \mathrm{K}$ \citep{magic2013a}, and so \avgNLTE\ NLTE analysis cannot be tested for HD222925. We thus warrant that a future full 3D, NLTE analysis is needed to fully investigate these trends, and suggest the adoption of 1D, NLTE abundances instead of 1D, LTE for now as they are more physically motivated.

\begin{figure}[ht]
\begin{center}
\vspace*{0cm}
\includegraphics[scale=0.8]{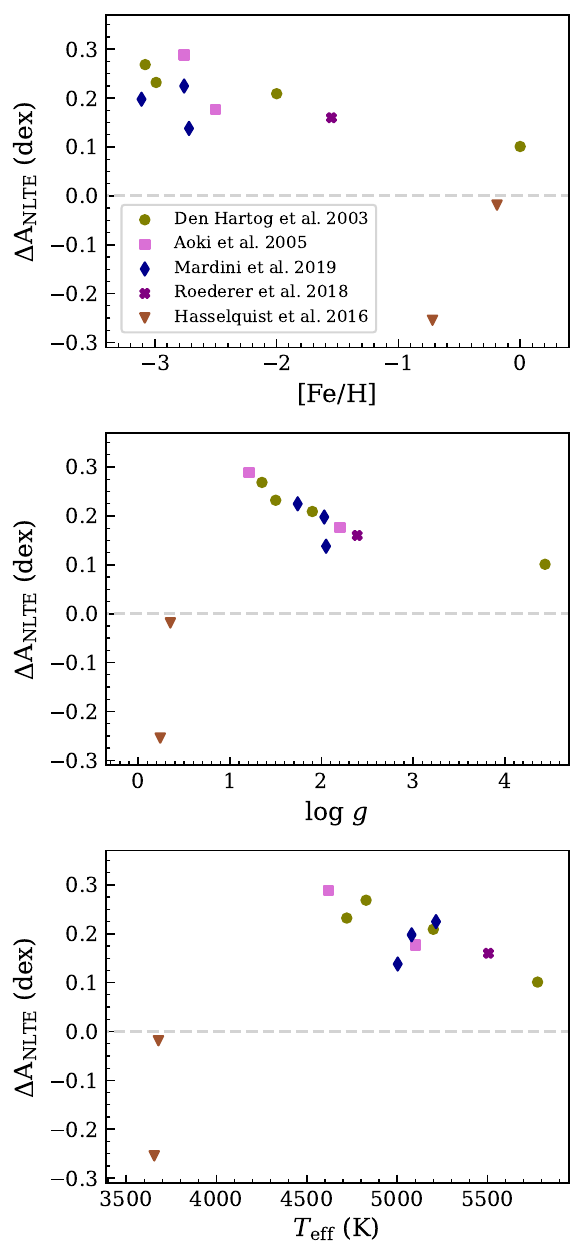}
\caption{\label{fig:nlte_sample} Plots of \dnlte\ versus \feh\ (top), \logg\ (middle), and \teff\ (bottom) for the stars in Table \ref{tab:star_list}. Each point represents the mean \dnlte\ value calculated using spectral line measurements from a single paper.}
\end{center}
\end{figure}

We compute \dnlte\ for each star in our sample as a function of stellar parameters to investigate the magnitude of NLTE effects, as shown in Figure \ref{fig:nlte_sample}. 
The S-type stars HD35155 and HD49368, with spectral lines measured in the H band from \citet{Hasselquist_2016},  have negative NLTE corrections, whereas the stars with spectral lines bluer than 5900\,{\AA} have positive NLTE corrections. Based on these measurements, \dnlte\ for optical lines appears to increase as \feh, \logg, and \teff\ decrease.
The discrepancy in the \dnlte\ values and trends between different wavelength ranges is evidence that the NLTE correction for Nd varies significantly depending not only on stellar parameters, but also spectral line properties. We thus perform calculations for a larger range of stellar parameters and spectral lines to thoroughly investigate trends in \dnlte\ versus various stellar and line parameters.

\subsection{NLTE Correction Grid}\label{sec:nlte_correction_grid}

We selected a parameter space typical of FGK-type metal-poor stars in which to calculate \dnlte\ corrections for 122 \ndii\ lines that have previously been measured in these stars. Our grid of stellar parameters covers the ranges $4000\ \mathrm{K} \le $ \teff\ $  \le 6500\ \mathrm{K}$ (step size 250 K), $0.50 \le $ \logg\ $ \le 4.50$ (step size 0.2\,dex), $-3.00 \le $ \feh\ $ \le -1.00$ (step size 0.25\,dex), and $1.0$ \kms\ $\le$ \vt\ $\le 2.0$ \kms\ (step size 1.0 \kms). Due to unreliable interpolation between atmospheric models, a small number of combinations of \logg\ (below 1.00\,dex) and \teff\ (above 6000 K) were omitted from the grid, but we do not expect many stars of interest to fall within this region of parameter space.

\begin{figure}[ht!]
\begin{center}
\vspace*{0cm}
\hspace*{-.3cm}
\includegraphics[scale=0.7]{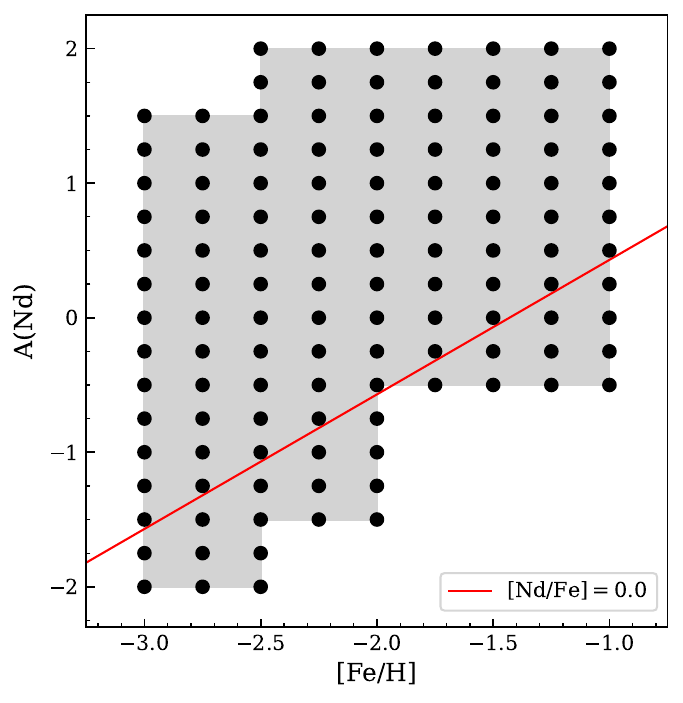}
\caption{\label{fig:logeps_grid}A grid displaying all values of \logepsLTE\ chosen for each metallicity value in our parameter space ($-3.0 \le$ \feh\ $\le -1.0$). A line showing \AB{Nd}{Fe} $= 0.0$ is provided for reference.}
\end{center}
\end{figure}

For each of these sets of stellar parameters, a generalized curve of growth (GCOG) was generated for each spectral line of interest using the method described in Section \ref{sec:testing_sun} with a higher order polynomial to fit a larger number of Nd abundances.
To avoid excessive computing time and convergence errors while still covering stars of interest for \rproc\ research, we selected a set of \logeps\ values for each metallicity as shown in Figure \ref{fig:logeps_grid}. The general parameter space covered by the grid was chosen by referencing literature data from JINAbase \citep{jinabase2018}.

In total, we ran \texttt{MULTI2.3} for 54,450 combinations of \teff, \logg, \feh, \vt, and \logeps.
For each of the sets of stellar parameters described above, we created GCOGs by fitting a degree 5 polynomial to the \texttt{MULTI2.3} EW output for between 11 and 17 \logeps\ values (shown in Figure \ref{fig:logeps_grid}), in both NLTE and LTE for each spectral line. 
Interpolation with a lower order polynomial over a smaller neighborhood of grid points would produce similar results, but as we demonstrate in the following section, our chosen method is sufficiently precise and accurate.
We then determined \dnlte\ from the abundance difference between the LTE and NLTE GCOGs at the LTE EW corresponding to each grid point.
To create a useful grid, we compiled a list of all \ndii\ lines that had been measured at least once in our sample of stars (see Section \ref{sec:metal_poor_star}), as well as five additional H band lines from \citet{Hasselquist_2016}. 
This results in a 6-dimensional grid containing \dnlte\ for each set of parameters \teff, \logg, \feh, \vt, \logepsLTE, and spectral line wavelength $\lambda$.

\subsection{Testing Grid Interpolation Accuracy}\label{sec:interpolation}

\begin{figure}[t!]
\begin{center}
\vspace*{0cm}
\hspace*{-.3cm}
\includegraphics[scale=0.75]{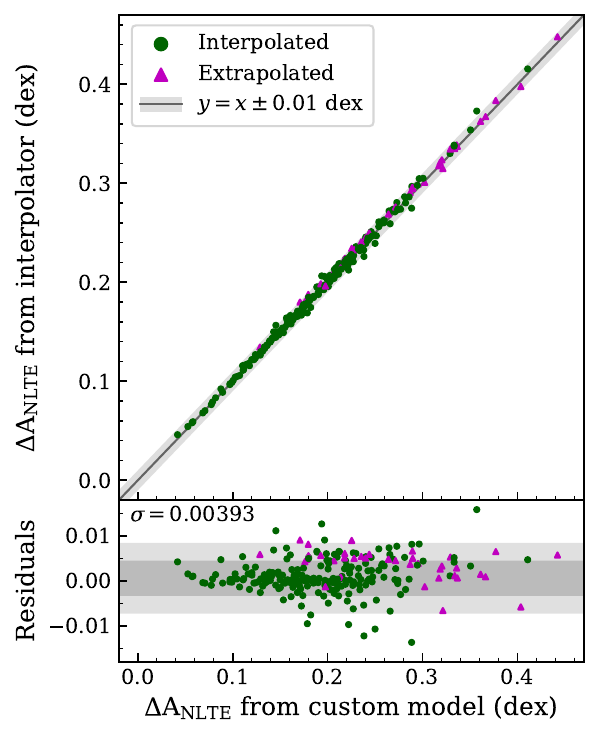}
\caption{\label{fig:correction_method_comparison}Comparison of \dnlte\ calculation methods for 248 measurements of optical \ndii\ lines in FGK stars. Interpolated points are shown as green circles, and extrapolated points are shown as purple triangles. A shaded region is included to highlight points that are within 0.01\,dex of exact agreement. Residuals from the one-to-one line are shown ($\mu = 0.00140$, $\sigma = 0.00393$) with shaded regions indicating $\pm1\sigma$ and $\pm2\sigma$.}
\end{center}
\end{figure}

Having generated a grid of evenly spaced \dnlte\ values within a large parameter space, we aim to demonstrate that linear interpolation between pre-calculated grid points produces sufficiently accurate NLTE corrections for the 122 selected \ndii\ spectral lines at most stellar parameter combinations. We thus interpolate NLTE corrections for each spectral line for 9 stars from Table \ref{tab:star_list}, excluding the two S-type stars (HD49368 and HD35155) and the Sun, as their parameters fall significantly outside of the grid. While some stars fall slightly outside the grid with [Fe/H] $\ge -3.11$ or \vt\ $\le 2.37$ \kms, their parameters are close enough to the grid range so we chose to extrapolate for their parameters. We then compare the interpolated NLTE corrections with the line-by-line derived NLTE corrections from Table \ref{tab:star_list}. The comparisons are shown in Figure \ref{fig:correction_method_comparison}.
The two methods produce nearly identical corrections for the spectral lines for all of the chosen stars, where the mean difference in \dnlte\ between the two methods is 0.001\,dex, with a standard deviation of 0.004\,dex. The largest difference is 0.016\,dex, and 232 of the 238 interpolated values ($97.5\%$) lie within 0.01\,dex of the one-to-one line. This gives us confidence that linear interpolation within the parameter space of our grid will produce sufficiently accurate corrections for NLTE analysis of metal-poor stars.

\section{Results and Discussion}\label{sec:results}

In this section, we present an analysis of our results for the \dnlte\ calculations described in the sections above, focusing on NLTE corrections as a function of both stellar parameters and spectral line properties, as well as discussing implications for \rproc\ enhanced stars and comparison to results from the literature.

\subsection{Variation of $\Delta \mathrm{A}_\mathrm{NLTE}$ with Stellar Parameters}\label{sec:variation_params}

\begin{figure*}[ht!]
\begin{center}
\vspace*{0cm}
\includegraphics[scale=0.51]{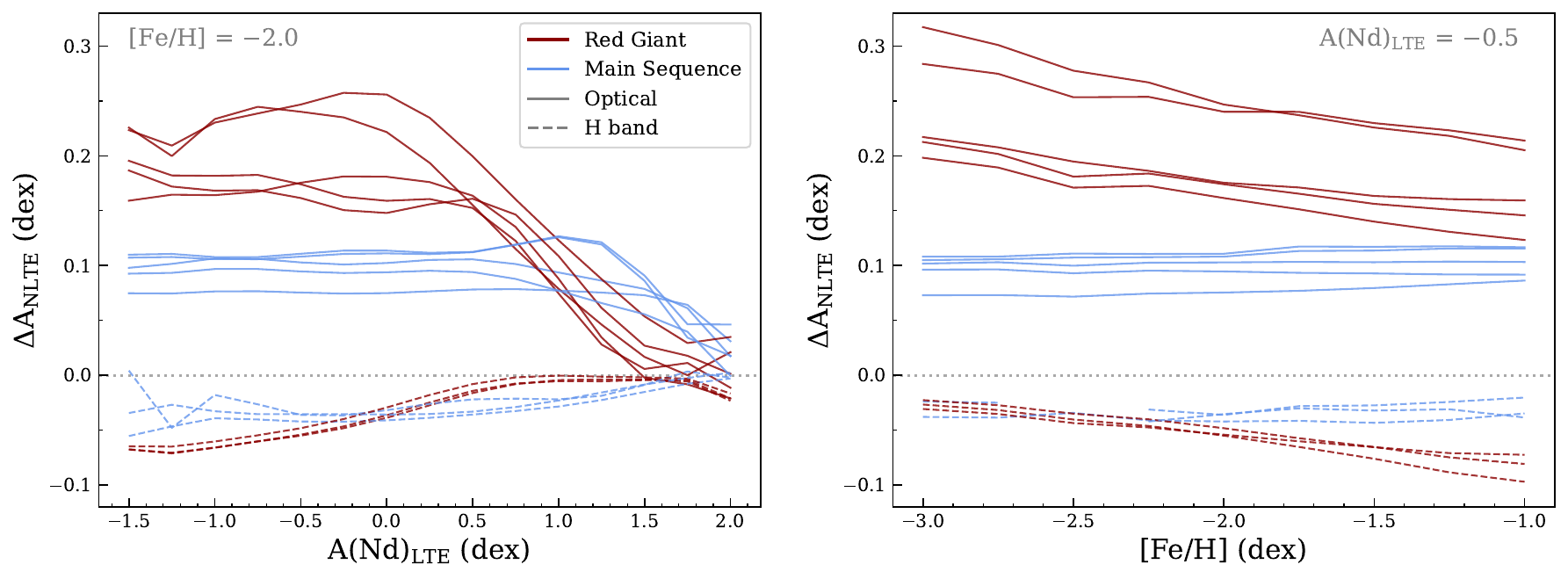}
\includegraphics[scale=0.51]{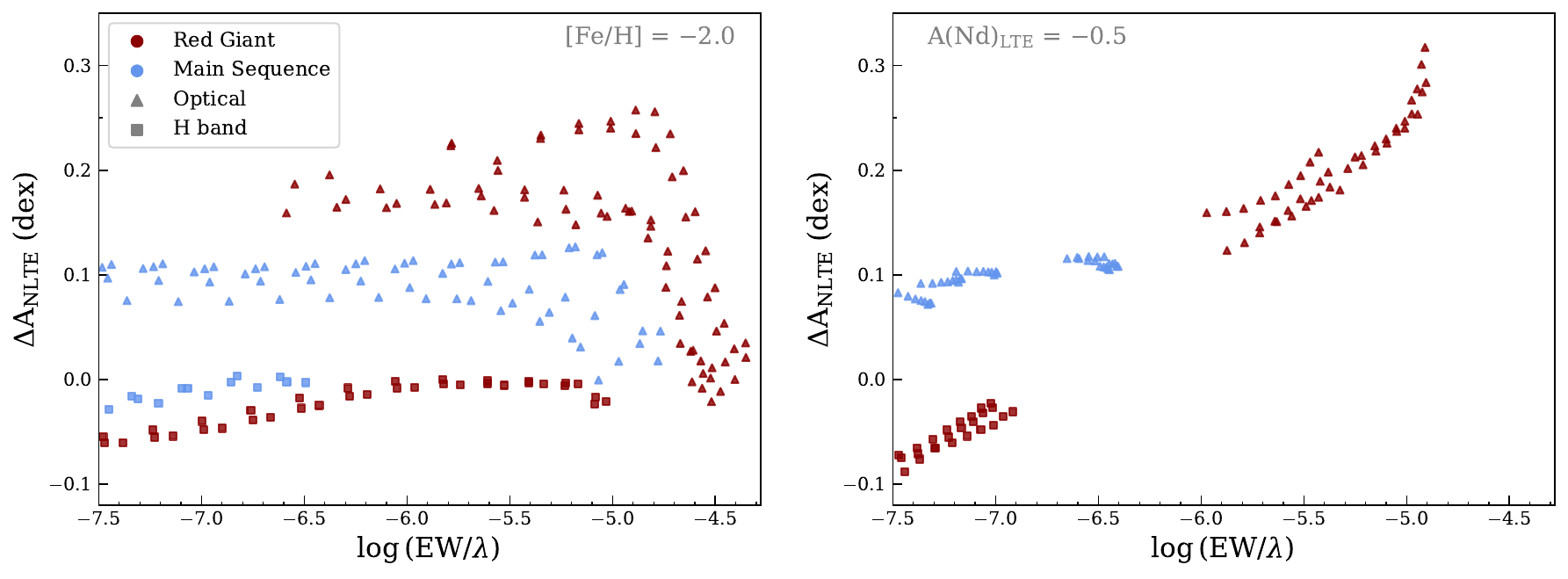}
\caption{\label{fig:dnlte_parameter_trends}NLTE corrections as a function of \logepsLTE\ (top left) and \feh\ (top right). Five commonly measured optical lines (4061.080, 4109.447, 4462.979, 5255.504, and 5319.813\,{\AA}) and three H band lines (15368.118, 16053.616, and 16262.042\,{\AA}) are shown in each graph. With the exception of the parameter being varied, dark red lines were calculated with typical red giant parameters (\teff\ $=4250 \mathrm{K}$, \logg\ $=1.50$, \feh\ $=-2.0$, \vt\ $=2.0$ \kms, \logeps\ $=-0.50$) and light blue lines were calculated with typical dwarf parameters (\teff\ $=5750 \mathrm{K}$, \logg\ $=4.30$, \feh\ $=-2.0$, \vt\ $=1.0$ \kms, \logepsLTE\ $=-0.50$). The bottom panels show the same data points as the top panels, but with REW plotted on the horizontal axis. A lower limit of $\log(\mathrm{EW}/\lambda) = -7.5$ is chosen to make the effects of saturation above $\log\mathrm{(EW/\lambda}) \gtrsim -5.0$ more visible.
}
\end{center}
\end{figure*}

With respect to the phase space of stellar parameters, NLTE corrections for \ndii\ appear to be highly degenerate and interdependent. Figure \ref{fig:dnlte_parameter_trends} shows NLTE abundance corrections calculated from GCOGs for a sample of models from our grid. By fixing stellar parameters to values corresponding to typical red giant (\logg\ $=1.5$) and dwarf main-sequence (\logg\ $=4.3$) stars, and plotting \dnlte\ versus \logepsLTE\ and \feh, respectively, for commonly measured optical and IR \ndii\ lines, we observe distinct trends for different types of stars and different lines. For optical lines in dwarf main-sequence stars, \dnlte\ is positive and relatively constant as \logepsLTE\ and \feh\ vary (with typical values of \dnlte\ $= 0.1$), only decreasing slightly to \dnlte\ $\simeq 0.05$ for highly Nd enhanced stars (\logepsLTE\ $\gtrsim 1.5$). NIR lines in the H band have a roughly constant negative correction of \dnlte\ $= -0.04$\,dex, although we note that NIR spectral lines at many main-sequence FGK parameters are far too weak to observe.

We find that for the model red giant star, \dnlte\ is typically larger than that for the model dwarf and depends on both \logepsLTE\ and \feh. Stars with \logeps\ $\lesssim 0.0$ have roughly constant NLTE corrections of $0.2$\,dex, while stars with \logeps\ $\gtrsim 0.0$ display a sharp decrease in NLTE corrections for the optical lines toward \dnlte\ $= 0$, as \logepsLTE\ increases. Corrections for the H band lines, on the other hand, slightly increase from $-0.07$ to $0.00$ dex as \logepsLTE\ increases. For a fixed value of \logepsLTE\ $= -0.50$, we find that NLTE corrections decrease from $0.30$ to $0.15$ dex for optical lines, and from $-0.02$ to $-0.08$ dex for H band lines, as \feh\ increases. 
The trends in \dnlte, particularly for giants, show a degeneracy with both metallicity and \logeps. As exemplified in the bottom panels of Figure \ref{fig:dnlte_parameter_trends} (particularly for optical lines in giants), for grid points where $\log\mathrm{(EW/\lambda}) \gtrsim -5.0$, there is a sharp change in the slope of \dnlte\ versus REW for each spectral line. Thus, the degeneracy is likely caused by saturation effects for \ndii\ lines with a REW above $-5.0$, as has been previously observed for other metals \citep{lind2011, reggiani2019, amarsi2022, lagae2025}. In this region of parameter space, we expect corrections to be less reliable due to differences in the COG slope and saturation point between NLTE and LTE models.

\begin{figure*}[ht!]
\begin{center}
\hspace*{-.2cm}
\par\bigskip
\includegraphics[scale=0.34]{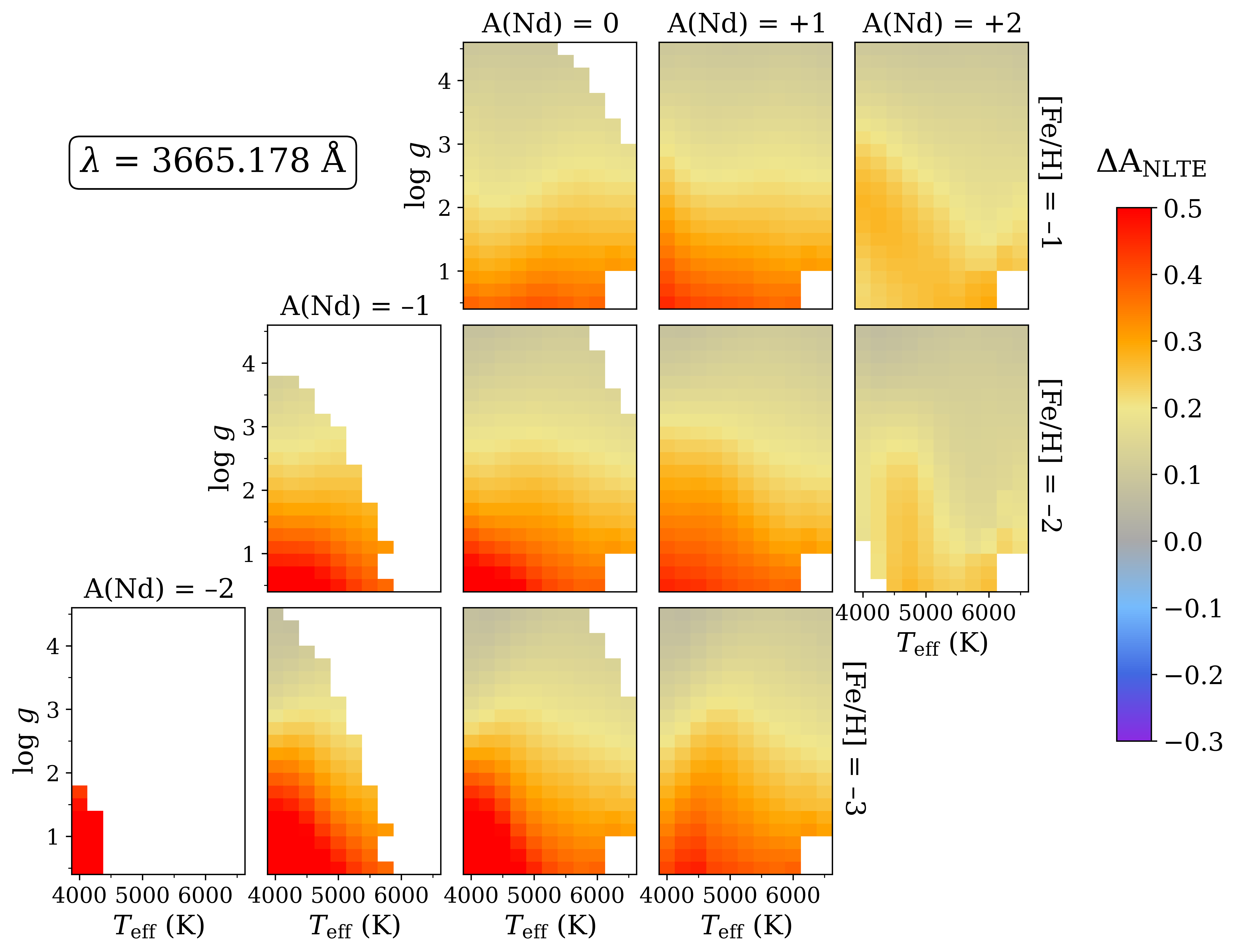}
\includegraphics[scale=0.34]{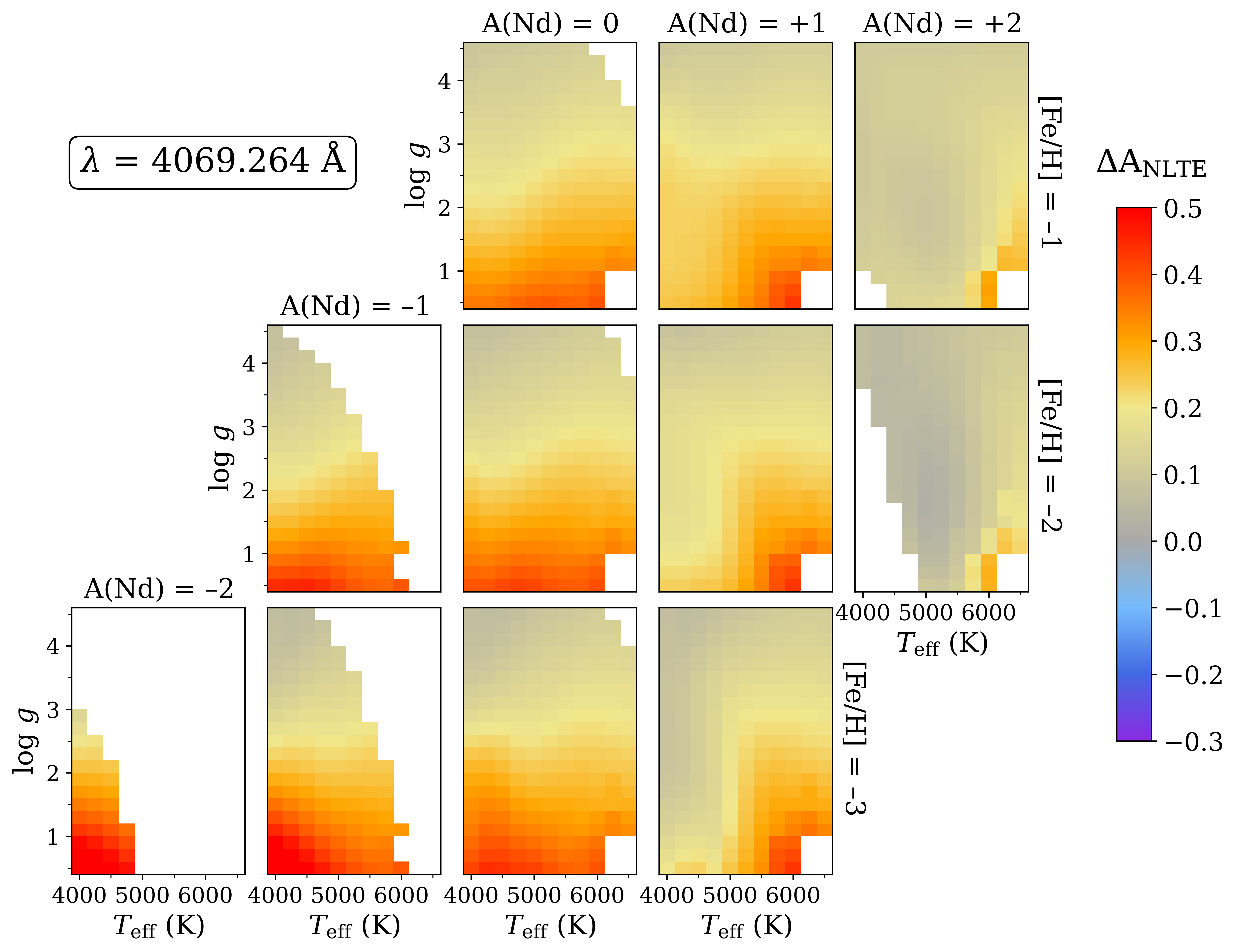}
\par\bigskip
\includegraphics[scale=0.34]{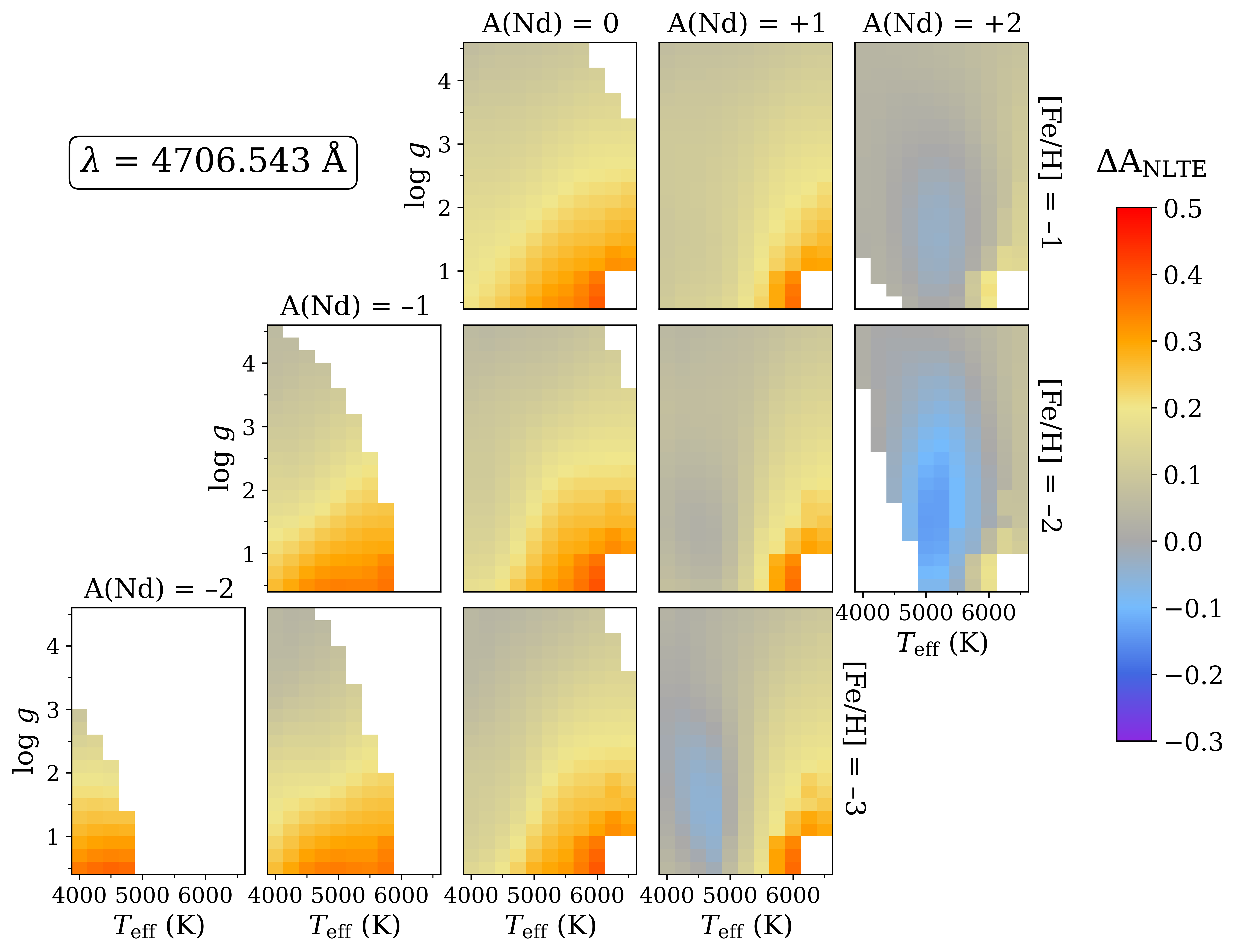}
\includegraphics[scale=0.34]{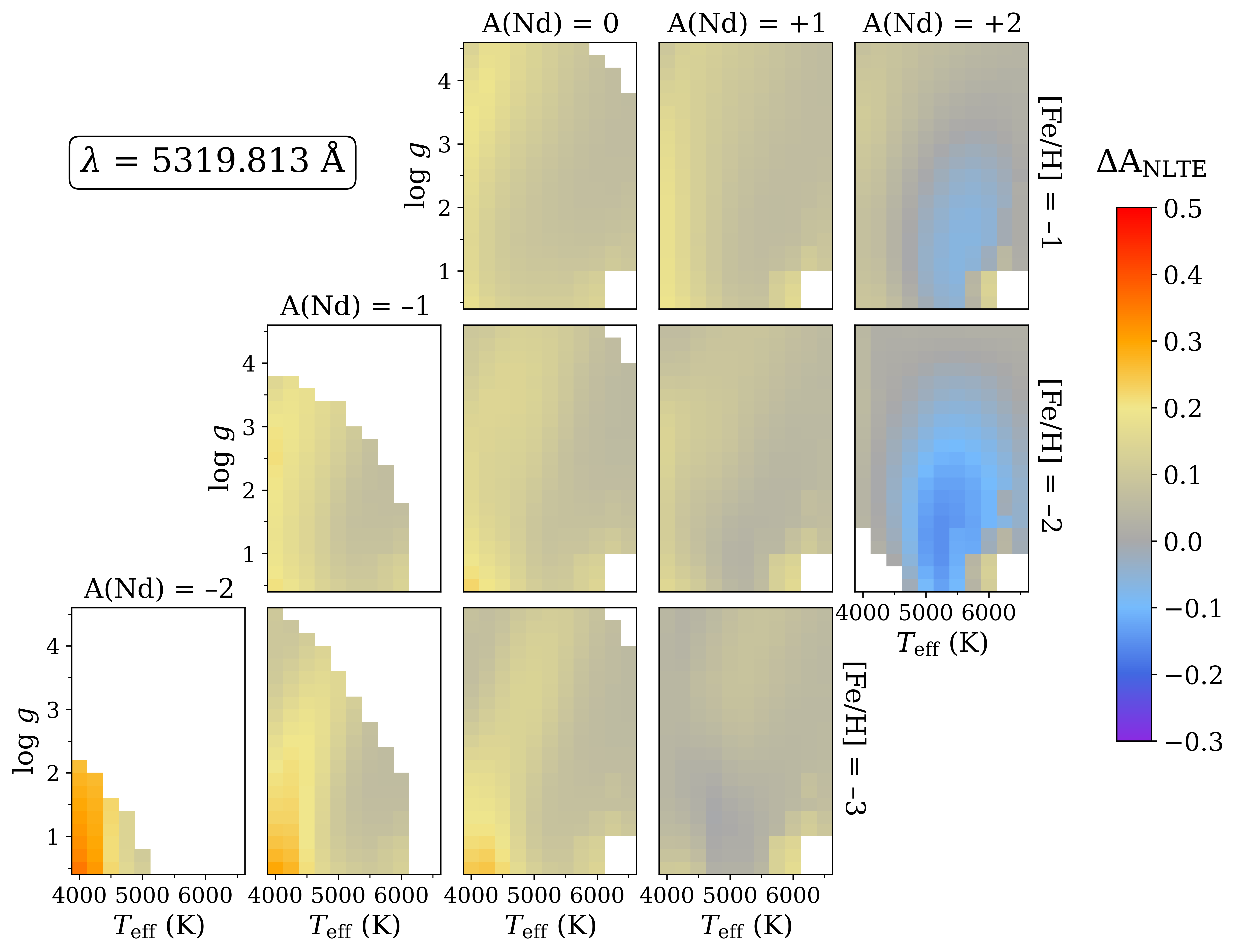}
\par\bigskip
\includegraphics[scale=0.34]{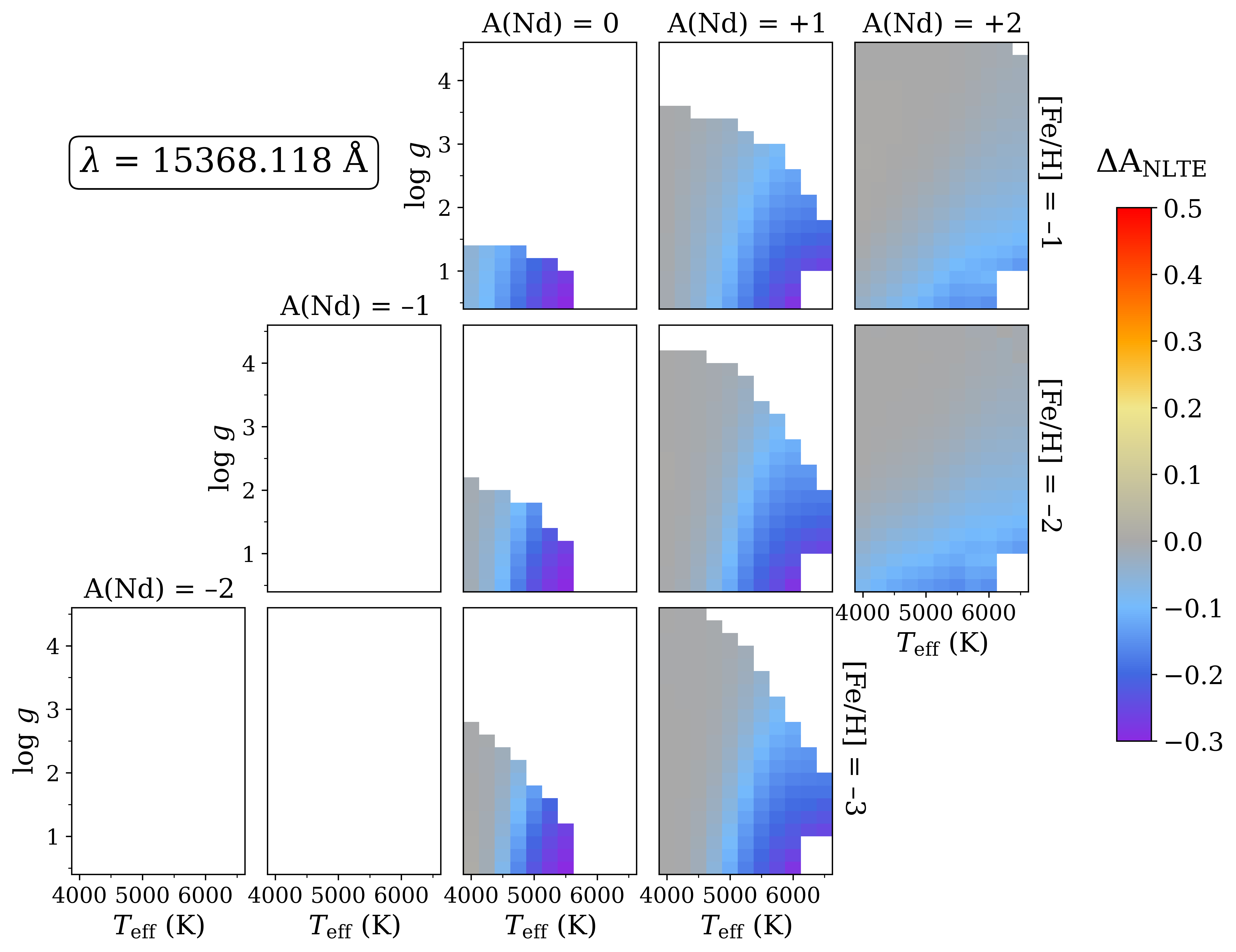}
\includegraphics[scale=0.34]{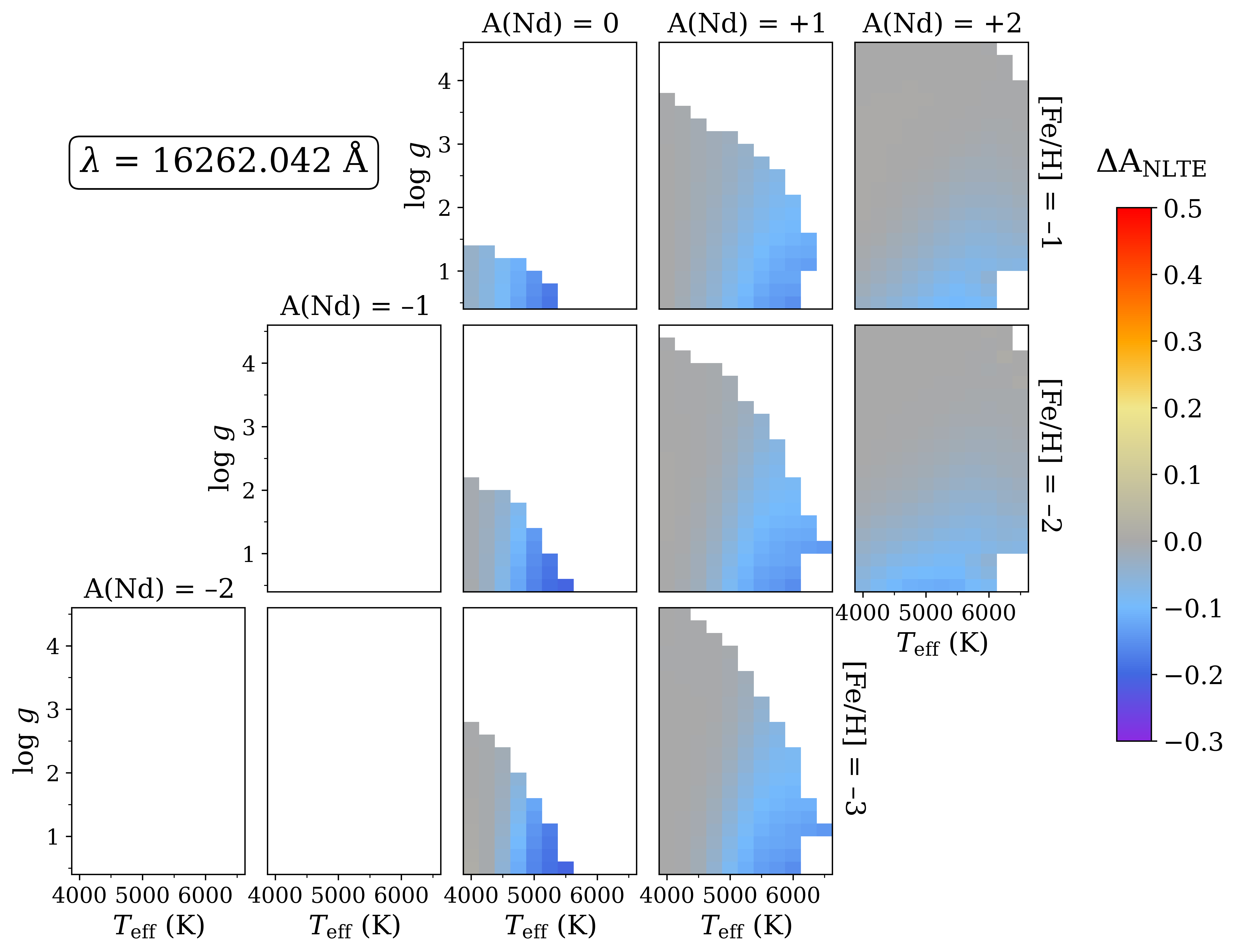}
\caption{\label{fig:heatmaps}Heat maps of \dnlte\ for six \ndii\ lines across the optical and NIR. Each individual panel in a heat map shows variations with \teff\ and \logg\ for a specific combination of \feh\ and \logepsLTE\ as indicated by the labels surrounding the panel. A value of \vt\ $=1.5$ \kms\ is interpolated for all heat maps. Corrections are omitted if the interpolated model atmosphere was unreliable, or if the REW does not fall in the range $-7.0 \le \log(W_\lambda / \lambda) \le -4.5$.}
\end{center}
\end{figure*}

To further demonstrate trends within the NLTE correction grid, we compute \dnlte\ corrections for each line in our sample of 122 \ndii\ lines, as a function of \teff, \logg, \feh, and \logeps, as shown in the heat maps of Figure\,\ref{fig:heatmaps} for a number of lines across a range of wavelengths in the optical and NIR. The plots only display \dnlte\ in the regions of parameter space where the REW lies between $-4.5$ and $-7.0$, which is a typical range used for analysis with the COG method.
As mentioned above, \ndii\ lines with a REW above $-5.0$ are not as reliable due to potential saturation issues. Our sample of observations from Section \ref{sec:nlte_analysis} does not appear to suffer from these issues, as \ndii\ lines are weak in most stars, and the small handful of strong lines observed in highly \rproc\ enhanced stars do not show unusual trends in \dnlte. However, we note this loose upper limit on line strength as a point of caution when performing NLTE abundance analysis of stars with high \rproc\ enrichment.

We find that among blue lines ($\lambda < 4500$\,{\AA}), NLTE corrections generally tend to be larger (up to +0.5\,dex) for cooler giants with lower \logg\ values and lower metallicities. Among those giants, the largest corrections are obtained for lower temperatures around 4000--4500 K. We also find that the corrections decrease as \logeps\ increases from $-2.0$ to $+2.0$, where they become close to 0 for most parameters.

For optical lines with  $4500 < \lambda <6000$\,{\AA}, we find that for stars with \logepsLTE\ $ < 1.0$, \dnlte\ increases up to 0.3 for warmer giant stars (\teff\ $>5000$\,K and \logg\ $< 2$), while remaining $<0.1$ for the rest of parameters. For \logeps\ $= +2.0$, \dnlte\ can decrease down to $-0.1$. For NIR lines in the H band ($\lambda > 14000$\,{\AA}), we find that corrections are negative, decreasing down to $-0.3$ for warmer (\teff\ $>5000$\,K) giants. 

\begin{figure*}[ht!]
\begin{center}
\vspace*{0cm}
\includegraphics[scale=0.68]{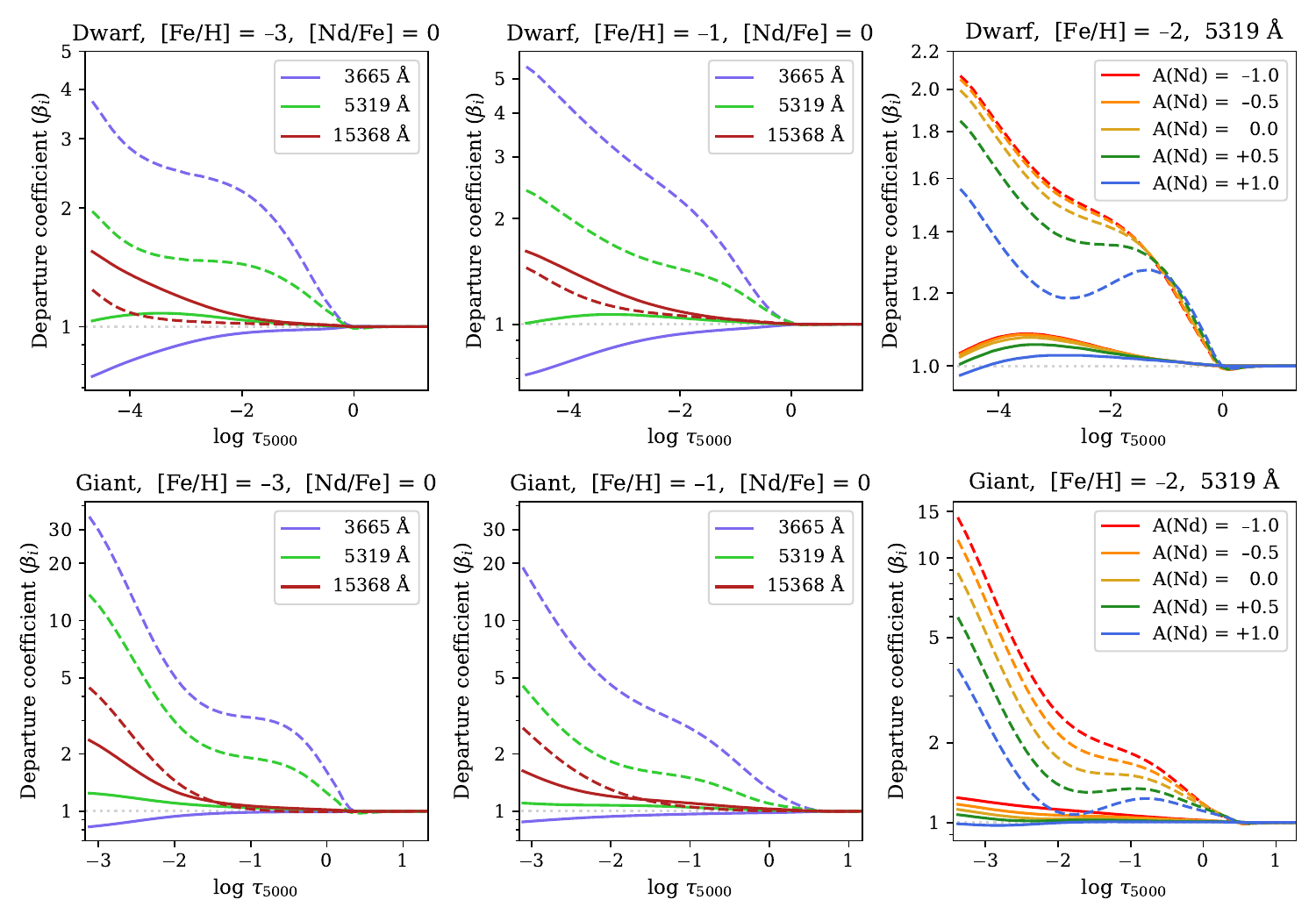}
\caption{\label{fig:optical_depth} Departure coefficients as a function of optical depth at 5000 \AA. The left and middle columns of panels show departure coefficients for the lower (solid) and upper (dashed) levels of a near-UV, optical, and NIR spectral line, for model atmospheres of a typical dwarf (\teff\ $= 5750\ \mathrm{K}$, \logg\ $= 4.3$, \vt\ $= 1.0\ \mathrm{km\ s}^{-1}$) and giant (\teff\ $= 4250\ \mathrm{K}$, \logg\ $= 1.5$, \vt\ $= 2.0\ \mathrm{km\ s}^{-1}$). The top right and bottom right panels show departure coefficients for a single optical line (5319 \AA, EP = 0.55 eV) in both models at five different values of \logeps.
}
\end{center}
\end{figure*}

To investigate these large variations in NLTE effects for different stars, we present plots of departure coefficients $\beta_i = n_{i,\mathrm{NLTE}} / n_{i,\mathrm{LTE}}$ (the ratio of the population of an energy state $i$ in NLTE versus LTE) as a function of optical depth. The top left, bottom left, top middle, and bottom middle panels of Figure \ref{fig:optical_depth} show $\beta_i$ values in four model stars for three spectral lines: $\lambda = 3665\ \mathrm{\AA}$ (EP = 0.205 $\mathrm{eV}$), $\lambda = 5319\ \mathrm{\AA}$ (EP = 0.550 $\mathrm{eV}$), and $\lambda = 15368\ \mathrm{\AA}$ (EP = 1.264 $\mathrm{eV}$), respectively. The four models represent a typical dwarf (\teff\ $=5750\ \mathrm{K}$, \logg\ $= 4.3$, \vt\ $=1\ \mathrm{km\ s}^{-1}$) and red giant (\teff\ $=4250\ \mathrm{K}$, \logg\ $= 1.5$, \vt\ $=2\ \mathrm{km\ s}^{-1}$) with \feh\ $= -3$ and $-1$.

As demonstrated in the top left and top middle panels of Figure \ref{fig:optical_depth}, for a typical dwarf star, the upper levels of lower-EP optical and near-UV lines are overpopulated compared to LTE, and much more so than the lower levels. Thus, for the near-UV and optical line, the source function is significantly larger than the Planck function ($J_\nu/B_\nu \simeq b_\mathrm{upper} / b_\mathrm{lower} > 1$), weakening the spectral line and explaining the slightly positive NLTE corrections, which compares well to studies of other dominant species \citep{bergemann2014b, storm2023}. In the outer stellar atmosphere, the ratio $b_\mathrm{upper} / b_\mathrm{lower}$ is larger for the near-UV line than the optical, accounting for the slightly larger near-UV corrections in dwarf stars (as compared to redder optical lines) shown in Figure \ref{fig:heatmaps}. Conversely for the NIR line in dwarf stars, departure coefficients for the lower level slightly exceed those of the upper level, which is consistent with the negligible and slightly negative NIR corrections calculated for dwarf star parameters in our grid. 

The bottom left and bottom middle panels of Figure \ref{fig:optical_depth} explain the large NLTE corrections we see in near-UV and optical lines in giants as compared to those in the H band. Compared to dwarf stars, the departure coefficients for the lower levels of each line are similar, but the overpopulation in the upper levels is much more extreme. The ratio of $b_\mathrm{upper} / b_\mathrm{lower}$ is significantly higher than in the dwarf, and so NLTE corrections will be positive and large. For the 3665 \AA\ line at $\log \tau_{5000} = -3.0$, the departure coefficient is $\beta = 16$ for \feh\ $=-1$ and $\beta = 30$ for \feh\ $=-3$. This also shows that the departure from LTE in giants has a large dependence on metallicity, with more metal-poor giants having significantly stronger NLTE effects in the optical and near-UV. NLTE corrections are insignificant in the NIR for the stars chosen in Figure \ref{fig:optical_depth}, but the effects of stellar parameters on \dnlte\ in the H band can be seen more clearly in the bottom left and bottom right heat maps in Figure \ref{fig:heatmaps}.

In the top right and bottom right panels of Figure \ref{fig:optical_depth}, we show departure coefficients of the 5319 \AA\ line at different \logeps\ values for both a typical dwarf and giant with \feh\ = $-2.0$. The difference in the departure coefficients between the upper and lower levels of the line explains why the calculated NLTE corrections are generally stronger for lower \logeps\ values regardless of the star type. This also appears to be true even for spectral lines with negative corrections; if departure coefficients for an energy level are less than 1, the star with the highest \logeps\ abundance will generally have the largest departure coefficients (closest to unity).

\subsection{Variation of $\Delta A_\text{NLTE}$ with Spectral Atomic Line Properties}\label{sec:variation_lines}

\begin{figure*}[ht!]
\begin{center}
\vspace*{0cm}
\hspace*{0cm}
\includegraphics[scale=0.43]{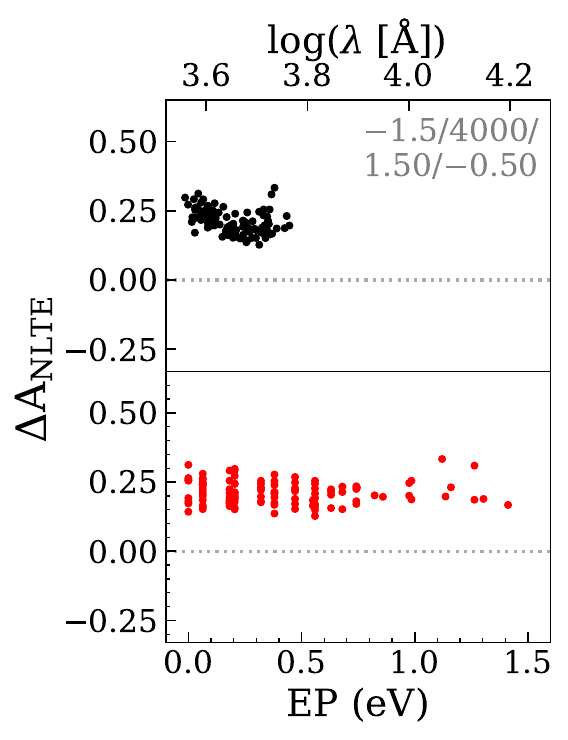}
\includegraphics[scale=0.43]{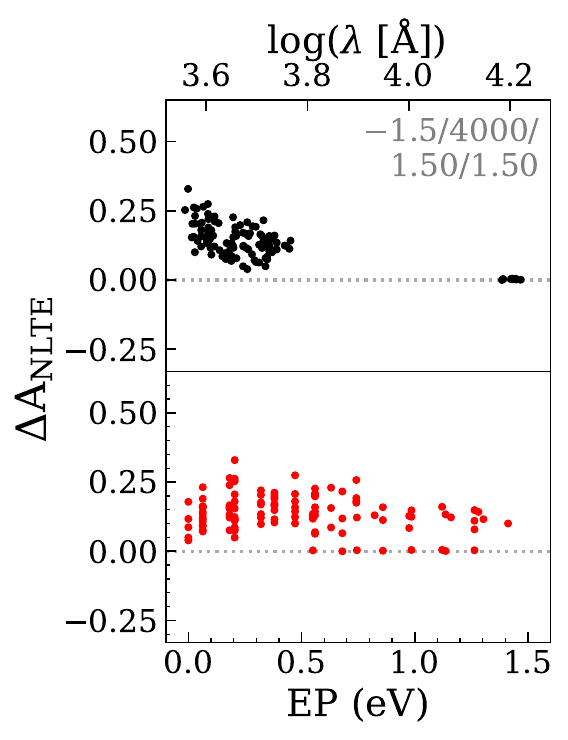}
\includegraphics[scale=0.43]{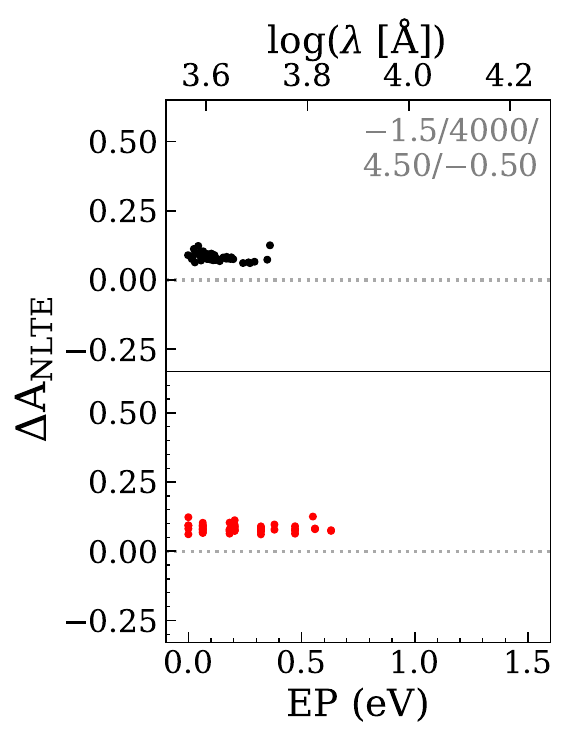}
\includegraphics[scale=0.43]{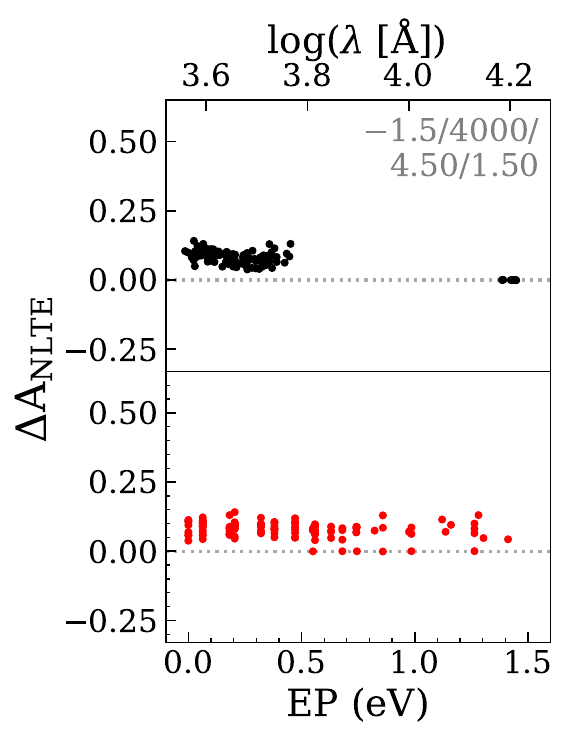}

\includegraphics[scale=0.43]{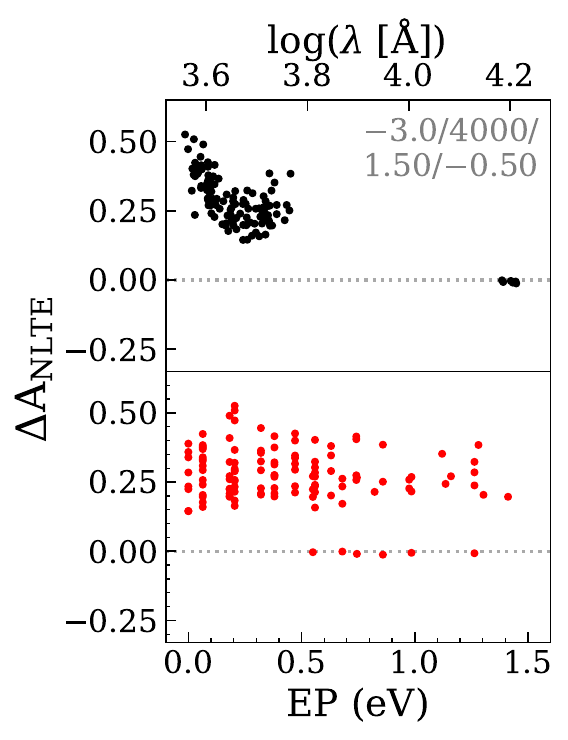}
\includegraphics[scale=0.43]{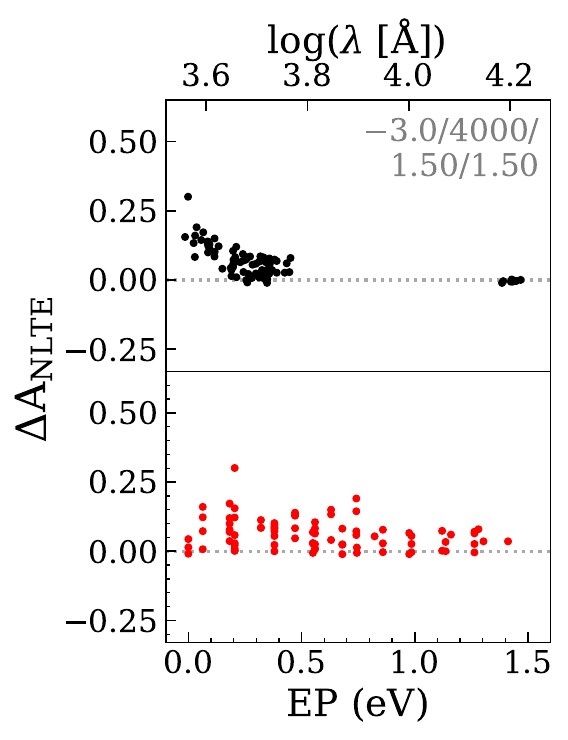}
\includegraphics[scale=0.43]{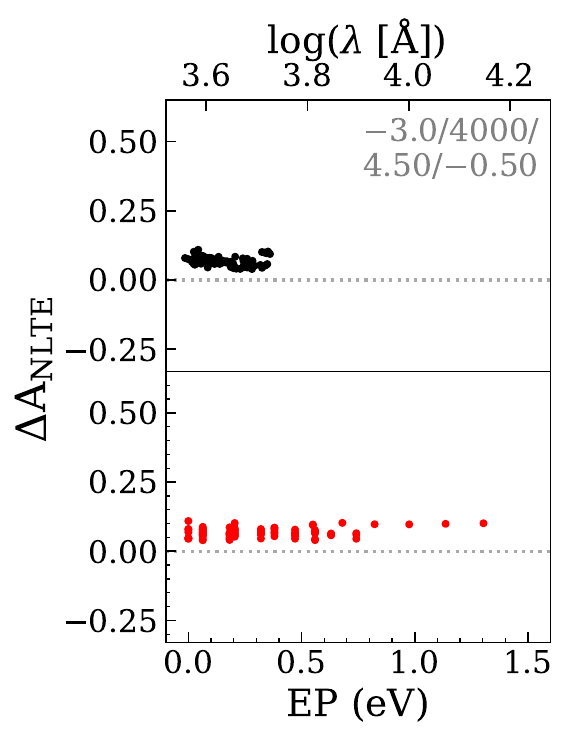}
\includegraphics[scale=0.43]{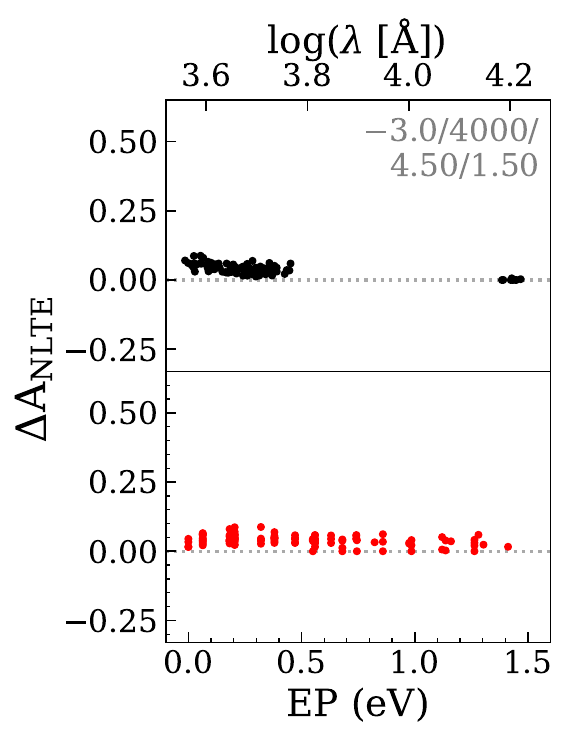}

\includegraphics[scale=0.43]{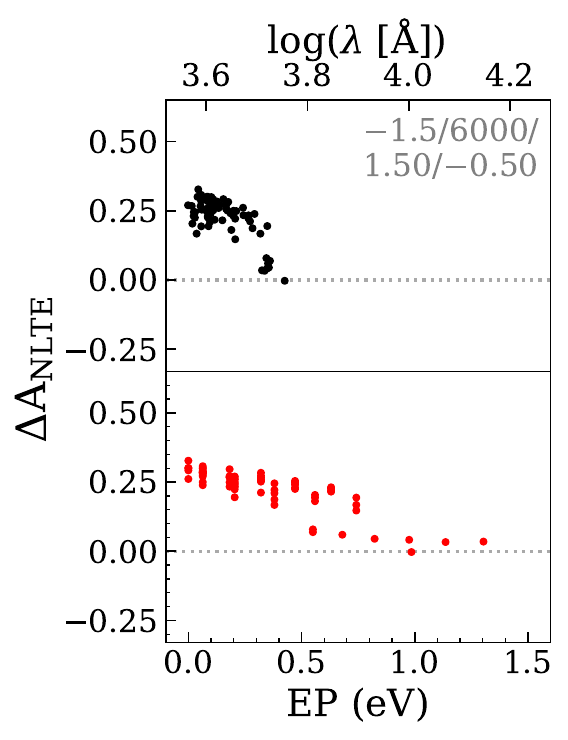}
\includegraphics[scale=0.43]{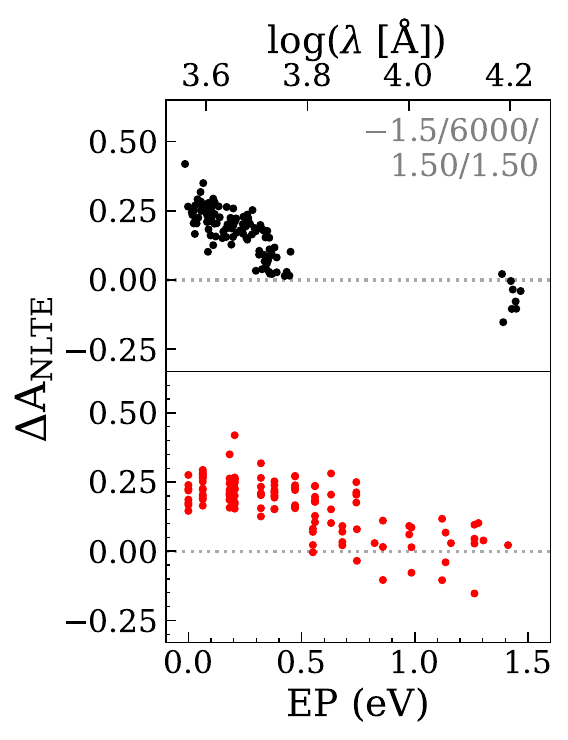}
\includegraphics[scale=0.43]{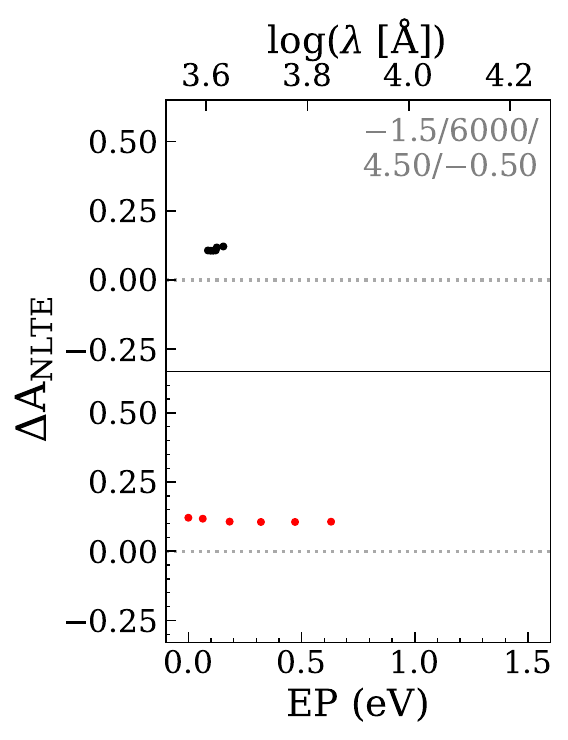}
\includegraphics[scale=0.43]{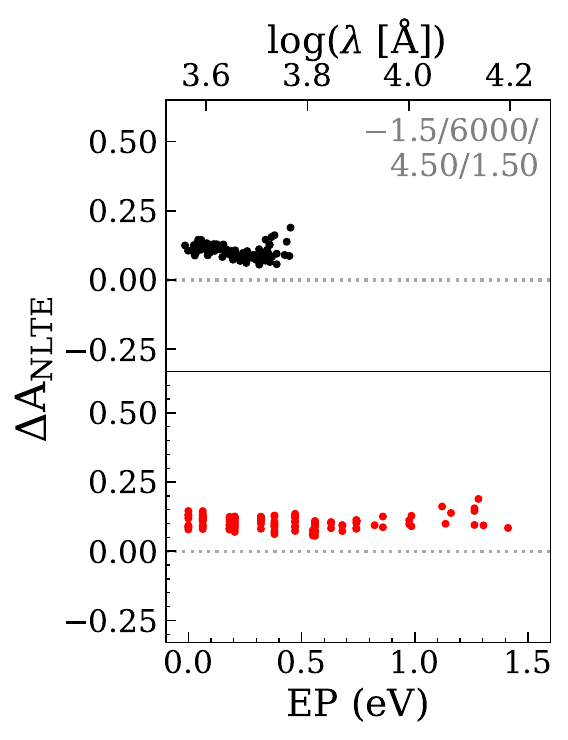}

\includegraphics[scale=0.43]{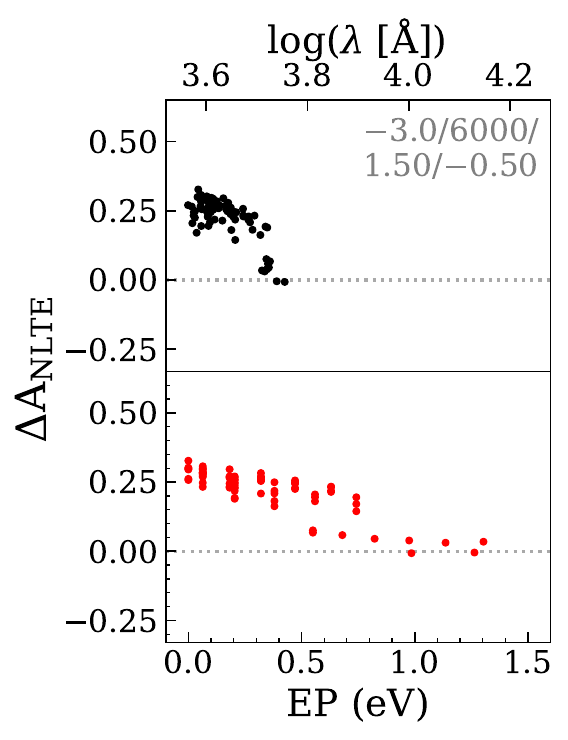}
\includegraphics[scale=0.43]{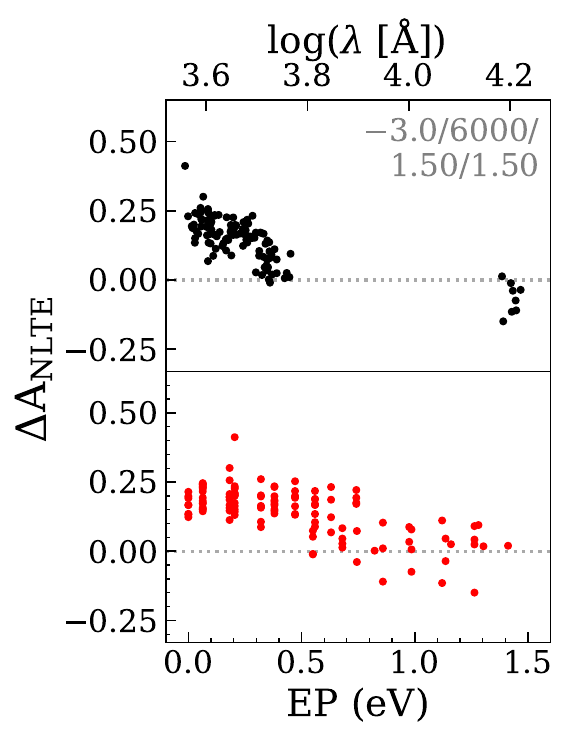}
\includegraphics[scale=0.43]{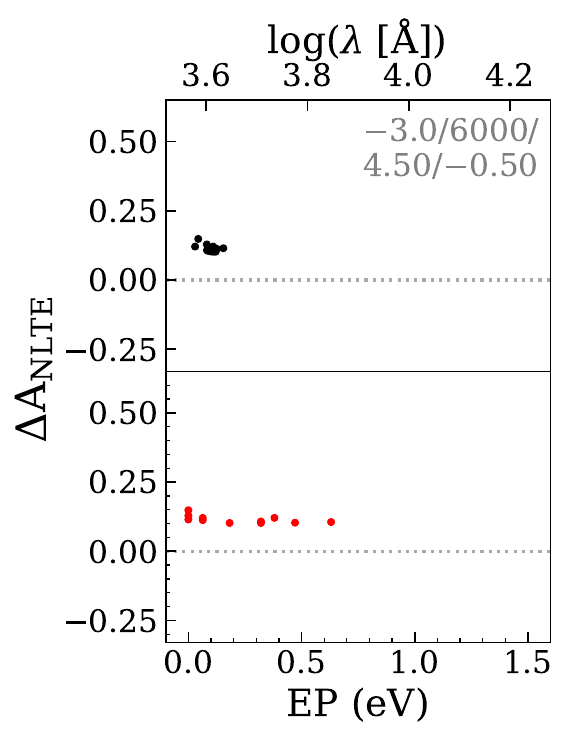}
\includegraphics[scale=0.43]{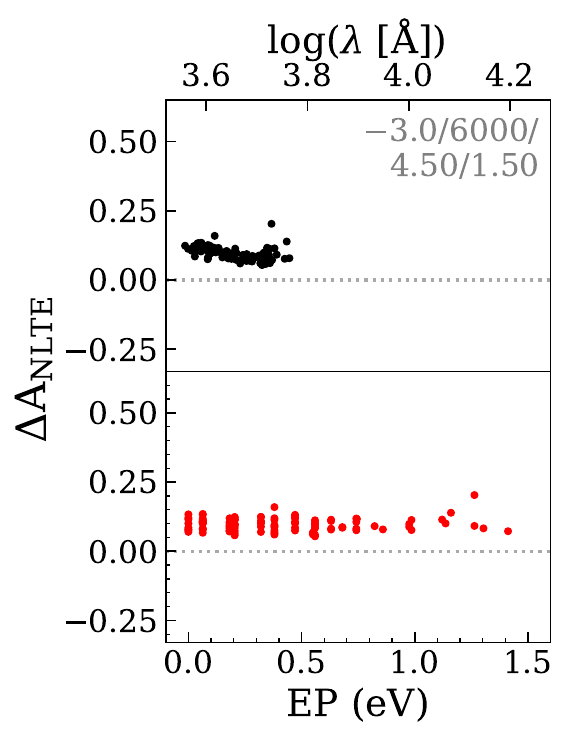}

\caption{\label{fig:wl_and_ep_plots}Plots of \dnlte\ as a function of both wavelength (top) and excitation potential (bottom) for 16 different combinations of stellar parameters \feh\ / \teff\ (K) / \logg\ / \logepsLTE\ near different extremes of the grid (indicated within each plot). We adopt \vt\ $=2.0$ \kms\ for giant stars (\logg\ $=1.50$) and \vt\ $=1.0$ \kms\ for dwarf stars (\logg\ $=4.50$). Hotter giant stars with the clearest negative trends are in the lower left quadrant. As with Figure \ref{fig:heatmaps}, corrections are omitted if $\log(W_\lambda / \lambda) < -7.0$ or  $\log(W_\lambda / \lambda) > -4.5$.}
\end{center}
\end{figure*}

As explained in the above sections, bluer near-UV \ndii\ lines tend to have overall more positive \dnlte\ values, while redder \ndii\ lines (including H band lines) have smaller and more negative corrections. To further investigate the dependence of the NLTE correction on spectral line properties, we
plot \dnlte\ versus both EP and wavelength for different sets of stellar parameters as shown in Figure \ref{fig:wl_and_ep_plots}. The plots show NLTE corrections for our sample of 122 \ndii\ lines, with  different combinations of \feh\ / \teff\ / \logg\ / \logepsLTE, for \feh\ $= -3.0$ and $-1.5$, \teff\ $=4000\,\mathrm{K}$ and $6000\,\mathrm{K}$, \logg\ $=1.50$ and $4.50$, and \logepsLTE\ $= -0.50$ and $1.50$. Similarly as above, only lines corresponding to a REW between $-7.0$ and $-4.5$ are included. 
The plots clearly display the trends with wavelength discussed in Section\,\ref{sec:nlte_analysis} for metal-poor giants, where \dnlte\ is highest for low EP and bluer lines, and decreases or remains roughly constant as EP increases. For dwarf stars, regardless of stellar parameters, \dnlte\ remains a constant ${\sim}0.1$\,dex as EP changes. The scatter in \dnlte\ as a function of EP is also much smaller for dwarfs than giants. In giant stars, corrections span a much wider range, from $-0.3$ to $0.5$ dex. Cooler giants have less of a downward trend with EP, and aside from NIR lines which have \dnlte\ $\simeq 0.0$\,dex, \dnlte\ values remain a constant $0.2$ or $0.3$ dex (depending on the star's metallicity and Nd abundance), with a large scatter. Hotter giant stars have the clearest downward trend, with low EP lines having corrections up to $0.3$\,dex and high EP lines having corrections as low as $-0.3$\,dex

We also observe that NLTE effects are generally weaker for higher Nd abundances, independent of changes in stellar parameters, as shown in the top right and bottom right panels of Figure \ref{fig:optical_depth}. Regardless if the NLTE corrections for a given star are positive or negative, they typically trend towards zero with increasing \logeps.

These NLTE effects are likely caused by a complex combination of physical processes. When $J_\nu > B_\nu$ in the UV, over-ionization occurs, which primarily causes the populations of lower levels to decrease, especially in stars with low \feh\ and \logg\ values. The analogous effect for bound-bound transitions is photon pumping, which can cause excess electrons from the lower level of a transition to populate the higher level if $J_\nu > B_\nu$ at the wavelength of the transition. Additionally, if $J_\nu < B_\nu$ in the infrared, a lack of ionization of higher EP transitions will cause over-recombination of the upper energy levels \citep{bergemann2014b}. The known effects of these processes on stellar abundance analysis generally align with the trends in \dnlte\ that we observe with respect to EP, particularly for metal-poor giants. However, it is difficult to fully predict how various physical processes interact within a stellar atmosphere, and a complete diagnostic analysis of the physical causes of these NLTE effects is outside the scope of this work.

As mentioned in Section \ref{sec:metal_poor_star}, a 3D, NLTE analysis may further help reduce the line-by-line trends seen as a function of EP and wavelength for many stars. In summary, we thus recommend the adoption of the 1D, NLTE abundance corrections to the commonly derived 1D, LTE abundances for r-process abundance analyses, particularly for metal-poor giants. We also recommend using moderate and high EP lines for abundance analysis of Nd whenever possible, as they will have smaller and more reliable NLTE corrections.

\subsection{Implications for \rproc\ stars}\label{sec}

Our NLTE corrections are comparable to previously computed NLTE corrections for dominant species of other neutron-capture elements. For example, a recent study by \citet{storm2023} found that NLTE corrections for \ion{Y}{ii} are as high as 0.4\,dex for low EP lines at parameters of \feh\ $\simeq -3.0$ and \logg\ $\simeq 1.0$. The corrections for  \ion{Y}{ii} follow the same general trends we obtained for \ndii, increasing for lower EPs, lower \logg\ values, and lower \feh\ values. We note that while the NLTE corrections can be larger than the typical systematic uncertainties, they are $<0.2$\,dex for most parts of the parameter space for both \ion{Y}{ii} and \ndii. 

Other recent papers find similar results for dominant species of common neutron-capture tracers. \citet{mashonkina2023} computes a grid of corrections for one \ion{Sr}{II} and \ion{Ba}{II} spectral line, finding that NLTE corrections range from $-0.2$ to $0.4$ dex, with large corrections derived for cool, metal-poor giant stars. For both of these lines, the NLTE corrections are also larger for lower abundances, matching our results for \ndii. \citet{guo2025} studied NLTE effects for two \ion{Eu}{II} lines at $4129\ \mathrm{\AA}$ and $6645\ \mathrm{\AA}$, computing positive corrections as high as $0.2$\,dex for the former and negative corrections as low as $-0.1$\,dex for the latter. For both lines, the deviation from LTE is stronger at lower metallicities, and giants show the most significant corrections. This study also applies these corrections to derived LTE abundances of both lines for a sample of metal-poor stars, finding that the two NLTE abundance values are in much closer agreement than the previous LTE values.

Based on all of these recent findings, it is possible that previously reported LTE abundances from optical lines of Nd (and possibly other neutron-capture elements) have underestimated the abundances of \rproc\ elements, particularly in giant metal-poor stars. While NLTE has been ignored for most previous studies, it is clear that NLTE calculations for each element must be performed for a wide range of parameters in order to assess the magnitudes of NLTE effects, even for dominant species like \ndii.

\section{Conclusions}\label{sec:conclusions}

We present NLTE abundance analysis of \ndii\ using a grid of 1D, NLTE corrections derived with the NLTE radiative transfer code \texttt{MULTI2.3}. We built a Nd model atom from up-to-date experimental \ndi\ and \ndii\ data and then calibrated the impact of hydrogen collisions using solar photospheric and meteoretic Nd abundances, determining \sh\ $=0.1$.
We incorporated theoretically computed \ndii\ transitions into our atom to assess the impact of atom completeness on our NLTE abundances, finding that the lack of experimental levels above 46000 cm$^{-1}$ has a negligible impact on the calculated NLTE corrections.

With this model atom, we generated synthetic spectra that successfully reproduced observations of \ndii\ spectral features for the Sun and the well-studied \rII\ star HD222925 \citep{roederer2018}, as well as a sample of Nd enhanced stars, for which we derived NLTE effects as large as $\pm 0.3$\,dex for multiple spectral lines in both the optical and H band. Most abundances in our sample were calculated from lines bluer than 5900\,{\AA}, and these abundances showed clear trends of \dnlte\ increasing as \teff, \logg, and \feh\ decrease. However, we find opposite effects for abundances calculated from H band lines, yielding negative NLTE corrections. To address the dependence of NLTE effects on changes in stellar atmospheric parameters, we generated a large grid of \dnlte\ corrections, covering parameters typical of FGK main-sequence and giant stars. 

We find significant trends in \dnlte\ within our grid, where corrections range from $-0.3$ to $+0.5$ dex depending on stellar parameters and spectral lines, which demonstrate the necessity of NLTE modeling for abundance analysis of \ndii. For a typical red giant, optical corrections are roughly ${\sim}0.2$\,dex for stars with lower Nd abundances but sharply decrease as the abundance rises above \logeps\ $\simeq 0.5$. For a typical G-type main-sequence star, optical NLTE corrections are typically ${\sim}0.1$, as evidenced by the abundance analysis of the Sun shown in Section \ref{sec:modelatom}. The largest positive corrections ($+0.5$\,dex) are for blue spectral lines ($\lambda < 4000$\,{\AA}) for the lowest values of \teff, \logg, and \feh\ in the grid. Negative corrections, as low as $-0.3$\,dex, are seen in NIR H band lines for combinations of low \logg\ and high \teff. Corrections generally tend to decrease as spectral line wavelength increases, and the trends in \dnlte\ across spectral lines show that lower EP lines tend to have larger corrections than higher EP lines, particularly in hotter giant stars. We anticipate the possibility that ignoring 3D effects could be contributing to the larger corrections for low-EP lines, and thus recommend adopting optical and NIR Nd lines with higher EPs for future \rproc\ abundance analysis when possible.

The non-negligible \dnlte\ values determined for \ndii\ and other neutron-capture species like \ion{Y}{ii} \citep{storm2023}, \ion{Eu}{ii} \citep{guo2025}, \ion{Sr}{ii} and \ion{Ba}{ii} \citep{mashonkina2023} at different stellar parameters are of high interest for \rproc\ research. These findings may have important implications for Galactic archaeology, as LTE analysis of blue lines in cool giants may significantly underestimate neutron-capture enhancement. Further investigations of NLTE effects for more neutron-capture species in \rproc\ enhanced stars, particularly lanthanides and elements around the second \rproc\ peak, are needed to assess the robustness of the corrections to the solar r-process pattern. This would allow more precise insights into the astrophysical conditions of the creation of these elements via the r-process.

Future efforts to expand this research will include continuing to improve the model atom as more experimental atomic data becomes available, as well as broadening the parameter range of our \dnlte\ grid to include spectral types like M and S. The current bounds of our grid were chosen to provide the most utility for \rproc\ research in the optical range, but it is clear from our results that NIR spectral lines in stars with \teff\ $< 4000\ \mathrm{K}$ require detailed NLTE calculations as well, especially as more IR spectra are released by SDSS-V, JWST, and other IR telescopes in the following years. 
We will also apply our NLTE corrections to a large sample of MW metal-poor stars to revisit the [Nd/Fe] trends compared to chemical evolution models, as well as investigate more NLTE corrections for other lanthanide \rproc\ elements (R. Shi et al., in prep.).

\begin{acknowledgments}

We are grateful to the anonymous referee for their valuable feedback and suggestions that have helped to improve this paper. The authors acknowledge UFIT Research Computing for providing computational resources and support that have contributed to the research results reported in this publication. URL: \href{http://www.rc.ufl.edu}{UFRC}.  

The authors would like to sincerely thank Dr.\@ Ian Roederer and Dr.\@ Jennifer Marshall for their valuable feedback and support during the research process. This research was completed with financial support provided by the Dr.\@ Dionel Avilés ’53 and Dr.\@ James Johnson ’67 Graduate Fellowship Program at Texas A\&M University. 
R.E. acknowledges support from JINA-CEE (Joint Institute for Nuclear Astrophysics Center for the Evolution of the Elements), funded by the NSF under Grant No. PHY-1430152., and support from NSF grant AST-2206263, and NASA Astrophysics Theory Program grant 80NSSC24K0899. This work also benefited from discussions at the 2024 Frontiers in Nuclear Astrophysics meeting at the University of Notre Dame and the Third Frontiers in Nuclear Astrophysics Summer School at Ohio University, supported by IReNA under NSF Grant OISE-1927130.

This work has made use of NASA's Astrophysics Data System Bibliographic Services; the  \href{https://arxiv.org/}{arXiv.org} preprint server operated by Cornell University; the SIMBAD and VizieR databases hosted by the Strasbourg Astronomical Data Center \citep{wenger2000}; the ASD hosted by NIST; and the VALD database operated at Uppsala University, the Institute of Astronomy RAS in Moscow, and the University of Vienna. Atomic data compiled in the DREAM database \citep{DREAM1} were extracted via VALD \citep[and references therein]{vald4}.

\end{acknowledgments}

\software{\texttt{matplotlib}~\citep{matplotlib},  \texttt{MULTI2.3}~\citep{Carlsson1986,carlsson1992}, 
\texttt{FORMATO3}~\citep{merle2011}, \texttt{Turbospectrum}~\citep{alvarez1998,plez2012}, TS-NLTE + \texttt{TSFitPy}~\citep{gerber2023,storm2023}, \texttt{LOTUS}~\citep{li2023}, \texttt{wrapper\_multi}~(\url{https://github.com/stormnick/wrapper_multi})}


\bibliography{ref}{}
\bibliographystyle{aasjournalv7}




\appendix
\onecolumngrid
\section{Additional Tables}
\restartappendixnumbering

\medskip

\begin{deluxetable*}{ccccccc}[bp]
\tablecaption{\label{tab:ews_list} Line-by-line Nd abundance values interpolated from derived COGs for each line, using the solar MARCS model atmosphere in both LTE and 4 different NLTE models with different $S_\text{H}$ values. Wavelengths and EWs are from VALD and \citet{denhartog2003}, respectively. Mean \logeps\ abundances with uncertainties are given at the bottom of the table.}
\tablehead{\colhead{$\lambda$ (\AA)} & \colhead{$W_\lambda$ (m\AA)} & \colhead{\logepsLTE} & \colhead{\logepsNLTE} & \colhead{\logepsNLTE} & \colhead{\logepsNLTE} & \colhead{\logepsNLTE} \vspace{-1.5mm} \\
{} & {} & {} & \colhead{$S_\text{H}=1.0$} & \colhead{$S_\text{H}=0.1$} & \colhead{$S_\text{H}=0.01$} & \colhead{$S_\text{H}=0.001$}} 

\startdata
4007.429 & 4.5 & 1.2898 & 1.3446 & 1.3798 & 1.3900 & 1.3913   \\
4021.326 & 11.4 & 1.2986 & 1.3751 & 1.4162 & 1.4282 & 1.4297  \\
4059.950 & 5.9 & 1.2734 & 1.3587 & 1.3926 & 1.4007 & 1.4017   \\
4136.745 & 1.3 & 1.2568 & 1.3495 & 1.3704 & 1.3742 & 1.3746   \\
4156.077 & 22.5 & 1.2867 & 1.3769 & 1.4227 & 1.4374 & 1.4394  \\
4284.509 & 5.5 & 1.2830 & 1.3491 & 1.3763 & 1.3823 & 1.3831   \\
4446.382 & 9.5 & 1.3002 & 1.3810 & 1.4169 & 1.4263 & 1.4274   \\
4465.592 & 2.1 & 1.3186 & 1.3803 & 1.4194 & 1.4326 & 1.4344   \\
4497.257 & 0.7 & 1.3916 & 1.4381 & 1.4729 & 1.4838 & 1.4849   \\
4567.604 & 1.3 & 1.3284 & 1.3985 & 1.4316 & 1.4408 & 1.4421   \\
4645.761 & 1.9 & 1.2886 & 1.3355 & 1.3619 & 1.3676 & 1.3682   \\
4715.585 & 4.0 & 1.4131 & 1.4803 & 1.5139 & 1.5240 & 1.5252   \\
4763.615 & 0.9 & 1.2836 & 1.3528 & 1.3788 & 1.3854 & 1.3863   \\
4777.716 & 1.4 & 1.4281 & 1.4704 & 1.5004 & 1.5086 & 1.5097   \\
4786.108 & 1.2 & 1.3546 & 1.4031 & 1.4451 & 1.4605 & 1.4626   \\
4797.153 & 2.6 & 1.3486 & 1.4278 & 1.4516 & 1.4547 & 1.4549   \\
4914.378 & 3.2 & 1.2689 & 1.3384 & 1.3644 & 1.3710 & 1.3717   \\
5063.722 & 1.3 & 1.3572 & 1.4528 & 1.4671 & 1.4605 & 1.4592   \\
5234.193 & 4.5 & 1.3823 & 1.4586 & 1.4665 & 1.4645 & 1.4642   \\
5255.504 & 5.9 & 1.3275 & 1.3913 & 1.4271 & 1.4377 & 1.4390   \\
5293.162 & 8.9 & 1.3641 & 1.4594 & 1.4683 & 1.4618 & 1.4607   \\
5306.457 & 0.9 & 1.4185 & 1.5363 & 1.5365 & 1.5284 & 1.5273   \\
5311.453 & 2.0 & 1.3464 & 1.4726 & 1.4784 & 1.4695 & 1.4681   \\
5319.813 & 8.7 & 1.3238 & 1.3982 & 1.4053 & 1.4027 & 1.4024   \\
5371.925 & 2.5 & 1.4377 & 1.5124 & 1.5241 & 1.5184 & 1.5174   \\
\\
\ Mean abundance & \  & 1.335 $\pm$ 0.053 & 1.410 $\pm$ 0.056 & 1.436 $\pm$ 0.050 & 1.440 $\pm$ 0.048 & 1.441 $\pm$ 0.048 \\
\enddata
\tablecomments{
More details on the source of information for each line are given in Table \ref{tab:140_lines}.}
\tablerefs{VALD, NIST, \citet{denhartog2003}}
\end{deluxetable*}

\begin{deluxetable*}{ccrcccrcccr}[bp]
\tablecaption{\label{tab:140_lines} Wavelenghths, excitation potentials, and $\log(gf)$ values for the 122 \ndii\ lines in the NLTE correction grid.}

\tablehead{\colhead{$\lambda$ (\AA)} & \colhead{EP (eV)} & \colhead{$\log(gf)$} & \colhead{\hphantom{XXXX}} & \colhead{$\lambda$ (\AA)} & \colhead{EP (eV)} & \colhead{$\log(gf)$} & \colhead{\hphantom{XXXX}} & \colhead{$\lambda$ (\AA)} & \colhead{EP (eV)} & \colhead{$\log(gf)$}}

\startdata
  3615.811$\ $ &   0.2046 &  $-0.760$\  &  &   4135.321$\ $ &   0.6305 &  $-0.070$\  &  &   4959.120$\ $ &   0.0636 &  $-0.800$\  \\
  3665.178$\ $ &   0.2046 &  $-0.660$\  &  &   4136.745$\ $ &   0.3802 &  $-1.030$\  &  &   4987.161$\ $ &   0.7421 &  $-0.790$\  \\ 
  3728.125$^a$ &   0.1823 &  $-0.500$\  &  &   4156.077$^a$ &   0.1823 &  $ 0.160$\  &  &   4989.419$^a$ &   0.6804 &  $-1.190$\  \\
  3738.056$\ $ &   0.5595 &  $-0.040$\  &  &   4177.319$^a$ &   0.0636 &  $-0.100$\  &  &   5063.722$\ $ &   0.9756 &  $-0.620$\  \\
  3759.794$^a$ &   0.6305 &  $-0.450$\  &  &   4211.289$\ $ &   0.2046 &  $-0.860$\  &  &   5066.828$\ $ &   0.5595 &  $-1.430$\  \\
  3763.471$^a$ &   0.2046 &  $-0.430$\  &  &   4232.374$\ $ &   0.0636 &  $-0.470$\  &  &   5092.792$\ $ &   0.3802 &  $-0.610$\  \\
  3780.382$^a$ &   0.4714 &  $-0.350$\  &  &   4284.509$^a$ &   0.6305 &  $-0.170$\  &  &   5130.585$^a$ &   1.3039 &  $ 0.450$\  \\
  3784.243$^a$ &   0.3802 &  $ 0.150$\  &  &   4303.571$^a$ &   0.0000 &  $ 0.080$\  &  &   5132.329$\ $ &   0.5595 &  $-0.710$\  \\
  3784.845$\ $ &   0.0636 &  $-1.040$\  &  &   4351.281$^a$ &   0.1823 &  $-0.610$\  &  &   5165.125$\ $ &   0.6804 &  $-0.740$\  \\
  3810.477$\ $ &   0.7421 &  $-0.140$\  &  &   4358.161$^a$ &   0.3206 &  $-0.160$\  &  &   5167.918$\ $ &   0.5595 &  $-1.180$\  \\
  3826.409$^a$ &   0.0636 &  $-0.410$\  &  &   4368.630$\ $ &   0.0636 &  $-0.810$\  &  &   5192.613$^a$ &   1.1365 &  $ 0.270$\  \\
  3838.980$\ $ &   0.0000 &  $-0.240$\  &  &   4385.660$^a$ &   0.2046 &  $-0.300$\  &  &   5212.358$\ $ &   0.2046 &  $-0.960$\  \\
  3879.541$\ $ &   0.3206 &  $-0.210$\  &  &   4400.821$^a$ &   0.0636 &  $-0.600$\  &  &   5215.651$\ $ &   1.2640 &  $-0.740$\  \\
  3887.867$\ $ &   0.0636 &  $-0.780$\  &  &   4446.382$\ $ &   0.2046 &  $-0.350$\  &  &   5234.193$\ $ &   0.5502 &  $-0.510$\  \\
  3890.937$\ $ &   0.0636 &  $-0.220$\  &  &   4451.978$\ $ &   0.0000 &  $-1.100$\  &  &   5249.574$^a$ &   0.9756 &  $ 0.200$\  \\
  3891.507$^a$ &   0.7421 &  $-0.140$\  &  &   4462.979$^a$ &   0.5595 &  $ 0.040$\  &  &   5250.812$^a$ &   0.7446 &  $-0.720$\  \\
  3900.218$^a$ &   0.4714 &  $ 0.100$\  &  &   4465.058$\ $ &   0.0000 &  $-1.360$\  &  &   5255.504$\ $ &   0.2046 &  $-0.670$\  \\
  3927.098$\ $ &   0.1823 &  $-0.590$\  &  &   4465.592$\ $ &   0.1823 &  $-1.100$\  &  &   5273.424$^a$ &   0.6804 &  $-0.180$\  \\
  3990.097$^a$ &   0.4714 &  $ 0.130$\  &  &   4497.257$\ $ &   0.4714 &  $-1.380$\  &  &   5293.162$^a$ &   0.8229 &  $ 0.100$\  \\
  3991.741$^a$ &   0.0000 &  $-0.260$\  &  &   4497.914$\ $ &   0.5595 &  $-1.020$\  &  &   5303.197$\ $ &   0.3802 &  $-1.430$\  \\
  3994.673$\ $ &   0.3206 &  $ 0.040$\  &  &   4501.809$\ $ &   0.2046 &  $-0.690$\  &  &   5306.457$\ $ &   0.8594 &  $-0.970$\  \\
  4004.004$\ $ &   0.0636 &  $-0.570$\  &  &   4506.582$\ $ &   0.0636 &  $-1.040$\  &  &   5311.453$^a$ &   0.9857 &  $-0.420$\  \\
  4007.429$\ $ &   0.4714 &  $-0.400$\  &  &   4541.267$\ $ &   0.3802 &  $-0.740$\  &  &   5319.813$^a$ &   0.5502 &  $-0.140$\  \\
  4011.063$\ $ &   0.4714 &  $-0.760$\  &  &   4542.600$\ $ &   0.7421 &  $-0.280$\  &  &   5356.965$^a$ &   1.2640 &  $-0.280$\  \\
  4012.243$^a$ &   0.6305 &  $ 0.810$\  &  &   4563.219$\ $ &   0.1823 &  $-0.880$\  &  &   5371.925$^a$ &   1.4124 &  $ 0.000$\  \\
  4012.697$\ $ &   0.0000 &  $-0.600$\  &  &   4567.604$\ $ &   0.2046 &  $-1.310$\  &  &   5431.514$\ $ &   1.1212 &  $-0.470$\  \\
  4013.216$\ $ &   0.1823 &  $-1.100$\  &  &   4645.761$\ $ &   0.5595 &  $-0.760$\  &  &   5485.097$\ $ &   0.5502 &  $-1.640$\  \\
  4018.822$\ $ &   0.0636 &  $-0.850$\  &  &   4703.571$\ $ &   0.3802 &  $-1.000$\  &  &   5485.693$^a$ &   1.2640 &  $-0.120$\  \\
  4021.326$\ $ &   0.3206 &  $-0.100$\  &  &   4706.543$\ $ &   0.0000 &  $-0.710$\  &  &   5688.518$^a$ &   0.9857 &  $-0.310$\  \\
  4023.005$\ $ &   0.5595 &  $ 0.040$\  &  &   4709.714$\ $ &   0.1823 &  $-0.970$\  &  &   5740.856$^a$ &   1.1602 &  $-0.530$\  \\
  4041.056$\ $ &   0.4714 &  $-0.530$\  &  &   4715.585$\ $ &   0.2046 &  $-0.900$\  &  &   5811.570$\ $ &   0.8594 &  $-0.860$\  \\
  4043.594$\ $ &   0.3206 &  $-0.710$\  &  &   4763.615$\ $ &   0.3802 &  $-1.270$\  &  &   5842.364$^a$ &   1.2816 &  $-0.600$\  \\
  4051.140$\ $ &   0.3802 &  $-0.300$\  &  &   4777.716$\ $ &   0.3802 &  $-1.220$\  &  &  15284.438$^b$ &   0.6804 &  $-2.130$\  \\
  4059.950$\ $ &   0.2046 &  $-0.520$\  &  &   4786.108$\ $ &   0.1823 &  $-1.410$\  &  &  15368.118$^b$ &   1.2640 &  $-1.550$\  \\
  4061.080$^a$ &   0.4714 &  $ 0.550$\  &  &   4797.153$\ $ &   0.5595 &  $-0.690$\  &  &  15912.275$^b$ &   0.5502 &  $-2.390$\  \\
  4069.264$\ $ &   0.0636 &  $-0.570$\  &  &   4799.419$\ $ &   0.0000 &  $-1.450$\  &  &  15977.939$^b$ &   1.1212 &  $-2.470$\  \\
  4075.111$\ $ &   0.2046 &  $-0.480$\  &  &   4820.337$\ $ &   0.2046 &  $-0.920$\  &  &  16053.616$^b$ &   0.7446 &  $-2.200$\  \\
  4075.272$^a$ &   0.0636 &  $-0.760$\  &  &   4825.476$\ $ &   0.1823 &  $-0.420$\  &  &  16262.042$^b$ &   0.9857 &  $-1.990$\  \\
  4109.070$\ $ &   0.0636 &  $-0.160$\  &  &   4859.025$\ $ &   0.3206 &  $-0.440$\  &  &  16303.772$^b$ &   0.8594 &  $-2.110$\  \\
  4109.447$^a$ &   0.3206 &  $ 0.350$\  &  &   4902.036$\ $ &   0.0636 &  $-1.340$\  &  &  16634.655$^b$ &   1.1365 &  $-2.370$\  \\
  4133.351$^a$ &   0.3206 &  $-0.490$\  &  &   4914.378$\ $ &   0.3802 &  $-0.700$\  &  &                &          &               
 \enddata
\tablecomments{If no footnote is present next to a wavelength, the wavelength and $\log(gf)$ value for that transition were acquired from VALD, but information on the upper and/or lower levels was missing and filled in by cross-referencing with the NIST ASD.}
\tablerefs{VALD, NIST ASD, \citet{Hasselquist_2016}}
\tablenotetext{}{\scriptsize $^{a\,}$All spectral line information for the transition was acquired from VALD. \\
$^{b\,}$Wavelengths, EPs, upper and lower energy levels, and log(gf) values acquired from \citet{Hasselquist_2016}.}
\end{deluxetable*}

\end{document}